\documentclass[a4paper, 10pt]{article}
\usepackage{graphicx}
\usepackage[usenames]{color}
\usepackage[flushleft]{threeparttable}
\DeclareGraphicsExtensions{.pdf,.png,.jpg,.mps,.eps,.ps}
\usepackage{amsmath,amssymb,gensymb}
\usepackage{natbib} 
\setcitestyle{authoryear,open={(},close={)}}
\usepackage{color}
\usepackage[toc]{appendix}
\usepackage[export]{adjustbox}
\usepackage{lscape}
\usepackage{enumitem}
\usepackage{psfrag}
\usepackage{graphicx,times}             
\usepackage{natbib}
\usepackage{amssymb,amsmath}
\usepackage{mathrsfs}
\bibpunct{(}{)}{;}{a}{}{,}

\usepackage{footnotehyper}

\usepackage{amsmath}
\usepackage{geometry}
\usepackage{graphicx,graphics}	% Including figure files
\usepackage[T1]{fontenc}
\usepackage{pbox}
\usepackage{booktabs}
\usepackage{tabularx}
\usepackage{setspace,hyperref}
\usepackage[utf8]{inputenc}
\usepackage{subcaption}

\def\apj{ApJ}
\def\apjs{ApJS}
\def\jcap{JCAP}
\def\mnras{MNRAS}

\def\nat{Nature}      
\def\apjs{ApJS}
\def\apjl{ApJ Letters}

\def\aaps{Astronomy and Astrophysics Supplement}

\title{Probing cosmic isotropy with Gamma-ray bursts: A dipole and quadrupole analysis of BATSE and Fermi GBM data}
\author{Debosi Mondal$^{1}$\thanks{E-mail: debosimondal@gmail.com}, Biswajit Pandey$^{1}$\thanks{E-mail: biswap@visva-bharati.ac.in (Corresponding author)}, and Amit Mondal\thanks{E-mail: amitmondal.bwn95@gmail.com}\\
  {\small$^{1}$ Department of Physics, Visva-Bharati University, Santiniketan, Birbhum, 731235, India }
  }

\date{\today}
\begin{document}
\pagestyle{empty}
\maketitle

\begin{abstract}
 
The cosmological principle, asserting large-scale homogeneity and isotropy, underpins the standard model of cosmology. Testing its validity using independent astronomical probes remains crucial for understanding the global structure of the Universe. We investigate the angular distribution of Gamma-Ray Bursts (GRBs) using two of the most comprehensive all-sky datasets available, the BATSE (CGRO) and Fermi GBM catalogs, to test the isotropy of the GRB sky at large angular scales. We perform spherical harmonic decomposition of the GRB sky maps and estimate the dipole and quadrupole amplitudes. Statistical significance is evaluated by comparing the observed multipole amplitudes against distributions derived from 500 Monte Carlo realizations of isotropic skies. Our results show that the observed dipole amplitudes for both BATSE and Fermi GBM datasets lie within the $1\sigma$ region of their respective null distributions. However, the quadrupole amplitude in the raw, uncorrected BATSE and Fermi GBM skies appears elevated at $3.7\sigma$ and $3.0\sigma$, respectively. After incorporating the BATSE sky exposure function, this apparent quadrupole anisotropy vanishes, indicating that instrumental non-uniformities fully account for the signal in that case. Owing to the absence of a publicly available full-sky exposure model for Fermi GBM, the Fermi analysis is restricted to the raw sky distribution. Our method’s reliability is validated through controlled simulations, which show it can detect the injected dipoles in BATSE-sized isotropic skies. These findings reinforce the statistical isotropy of the GRB sky and underscore the importance of accurate exposure corrections in cosmological anisotropy analyses.

\end{abstract}

%\begin{keywords}

\section{Introduction}
One of the most fundamental assumption in cosmology is the cosmological principle, the notion that, when viewed on large enough scales, the Universe exhibits no preferred location or direction. This principle forms the backbone of the Friedmann-Lemaitre-Robertson-Walker (FLRW) metric, the geometry behind our standard model of cosmology. A range of observations support this framework, with one of the most compelling evidences being the high degree of uniformity observed in the cosmic microwave background (CMB), a relic radiation from the early Universe \citep{penzias65, smoot96}. Additional evidence for isotropy comes from various independent tracers, including the angular distributions of gamma-ray bursts \citep{meegan92}, Type Ia supernovae \citep{gupta10}, distant radio sources \citep{blake02}, the X-ray background \citep{wu99}, large-scale galaxy surveys \citep{gibelyou12, pandey17b, sarkar19a, sarkar19b, camila24}, and the spatial distribution of galaxy clusters \citep{bengaly17}. Collectively, these observations provide broad support for the large-scale statistical isotropy of the Universe.

In recent years, several observational studies have investigated potential deviations from large-scale isotropy \citep{aluri22}. These include analyses of peculiar velocity flows in Type Ia supernovae \citep{colin19}, measurements of dipole anisotropies in the quasar distribution \citep{secrest21, secrest22, kothari22}, and reported hemispherical asymmetries in galaxy number counts \citep{wiegand14, appleby22}. Additionally, large-scale cosmic structures such as filaments and sheets \citep{sarkar23}, the Large Quasar Groups (LQGs) \citep{clowes13}, correlations in quasar orientations \citep{friday22}, and extensive underdense regions or voids \citep{keenan13, hasl20} have been identified and examined for their implications on isotropy. While each of these findings is accompanied by methodological uncertainties and remains under active investigation, they collectively underscore the importance of continued and more precise testing of the cosmological principle. Should statistically significant deviations from isotropy be established, it could prompt a re-evaluation of the standard cosmological framework based on the FLRW metric.

Gamma-ray bursts (GRBs) are among the most energetic and luminous events in the Universe, releasing intense bursts of gamma radiation that can outshine entire galaxies for brief periods. They are typically classified into long and short bursts, believed to originate from different astrophysical progenitors such as collapsing massive stars and compact binary mergers \citep{kouveliotou93}. Due to their brightness and detectability across cosmological distances, GRBs serve as powerful tracers of the large-scale structure of the Universe, making them promising probes for testing the fundamental assumption of cosmic isotropy.

A growing body of research has investigated the isotropy of the Universe using gamma-ray bursts (GRBs), offering valuable insights into the validity of the cosmological principle. The isotropic angular distribution of GRBs was among the earliest indications of their extragalactic origin, as highlighted in the seminal work by \citet{meegan92}, which showed that while the burst intensity distribution deviated from the expected Euclidean $-3/2$ power-law, the angular distribution remained statistically isotropic strongly disfavoring a Galactic origin. Building on this, \citet{briggs96} analyzed the first 1005 GRBs from BATSE using dipole and quadrupole statistics, concluding that the observed distributions were highly consistent with isotropy and far more uniform than any known Galactic population, thereby favouring a cosmological origin.

Subsequent investigations have uncovered a more nuanced picture, particularly when GRBs are split by duration or brightness. \citet{tarnopolski17} applied a suite of statistical tools to the Fermi/GBM dataset and found that while long GRBs appear isotropic, short GRBs display significant anisotropies with confidence levels exceeding $99.98\%$. Similarly, \citet{vavrek08} divided the BATSE sample into short, long, and intermediate-duration subgroups and applied a number of statistical tools including Voronoi tessellation, minimal spanning tree, and multifractal analysis concluding that short and intermediate GRBs deviate significantly from randomness, while long GRBs are consistent with an isotropic sky. Complementary results were reported by \citet{bernui08}, who used coordinate-independent angular correlation analyses of BATSE short GRBs and, after correcting for non-uniform sky exposure, found no intrinsic anisotropies, suggesting that observed deviations could arise from instrumental effects rather than cosmological ones. Further studies expanded this investigation beyond angular positions to redshift-dependent clustering. \citet{balazas99} identified a statistically significant quadrupole component in the BATSE sky map and suggested that the anisotropy observed particularly among short GRBs was not fully attributable to instrumental bias. \citet{mesazaros19} surveyed the body of work suggesting anisotropic patterns in GRB distributions and posited that the existence of giant structures traced by GRBs challenges the foundational assumptions of isotropy. \citet{ripa19} examined isotropy in GRB properties such as duration and fluence across several catalogs, and found broad consistency with statistical isotropy. \citet{andrade19} showed that position uncertainties in Fermi GBM GRBs can introduce spurious anisotropies in angular correlation functions, and once properly accounted for, the sky distribution of GRBs remains statistically isotropic even for short bursts. \citet{balazas18} found several ring-like spatial configurations in GRB redshift distributions, such as the Giant GRB Ring, but emphasized that such local irregularities do not justify rejecting the cosmological principle. \citet{bagoly22} similarly tested the spatial two-point correlation function using all known GRBs with spectroscopic redshifts and found no statistically significant large-scale clustering, apart from a marginal deviation in one sky patch. Most recently, \citet{horvath24} identified clustering in both Galactic hemispheres, including the Hercules-Corona Borealis Great Wall and the Giant GRB Ring, which have grown more prominent with the increasing GRB sample size. Though provocative, the interpretation of these structures remains contentious. Collectively, these studies demonstrate both the promise and the complexity of using GRBs to test cosmological isotropy, motivating the need for rigorous, simulation-based analyses to robustly assess potential deviations from the standard cosmological model.

The BATSE (Burst and Transient Source Experiment) catalog, compiled during the Compton Gamma Ray Observatory mission (1991–2000), includes 2702 GRBs with nearly full-sky coverage \citep{paciesas99}. The Fermi Gamma-ray Burst Monitor (GBM), operational since 2008 aboard the Fermi satellite, has detected over 4000 GRBs to date, with a broader energy range and improved temporal resolution \citep{vonkienlin20}. These two datasets represent the most extensive all-sky samples of GRBs available, enabling robust statistical analyses across different missions and detection strategies. Their combined use not only enhances the statistical power of our study but also allows for a cross-validation of isotropy tests across independent datasets.

A non-zero dipole in the distribution of GRBs would indicate a directional anisotropy, a statistically significant excess of GRBs in one region of the sky compared to its opposite. Similarly, a non-zero quadrupole in the distribution of GRBs would indicate a planar or axial anisotropy, suggesting a systematic variation in the number of bursts across two opposite hemispheres or along a preferred plane in the sky. These would challenge the cosmological principle, which assumes the Universe is isotropic on large scales.

In this study, we examine the angular isotropy of gamma-ray bursts using a statistical framework applied to two of the most comprehensive and widely used GRB datasets: the BATSE and Fermi GBM catalogs. We consider the combined GRB sample without separating long- and short-duration bursts. This approach implicitly assumes that, at the largest angular scales probed here, the combined GRB population provides a meaningful tracer of the underlying sky distribution. However, the long-duration and short-duration GRBs arise from different astrophysical progenitors, tracing different cosmic environments and evolutionary histories. The implications of this assumption for the interpretation of our results are discussed in the conclusions section. We measure the dipoles and quadrupoles in these datasets and assess their statistical significance. While previous analyses have considered dipole patterns in GRB distributions \citep{briggs96}, our methodology advances this line of inquiry by generating empirical dipole amplitude distributions from Monte Carlo simulations of isotropic skies. By comparing the observed dipole amplitudes to these simulated distributions, we obtain a direct and statistically robust estimate of the significance of any detected anisotropy. To further validate the reliability of our approach, we perform control tests using mock GRB skies with injected dipole signatures, thereby ensuring that our framework is sensitive to known deviations from isotropy. This simulation-driven methodology provides a systematic and data-driven means of testing the null hypothesis of angular isotropy in GRB distributions.

The remainder of this paper is organized as follows. In Section~\ref{sec:data_method}, we outline the datasets used and the methodology adopted. Section~\ref{sec:results} details the results of our analysis. Finally, Section~\ref{sec:conclusion} provides a summary of our findings and concluding remarks.

%%%%%%%%%%%%%%%%%%%%%%%%%%%%%%%%%%%%%%%%%

\begin{figure*}[htbp!]
\centering \includegraphics[width=12cm]{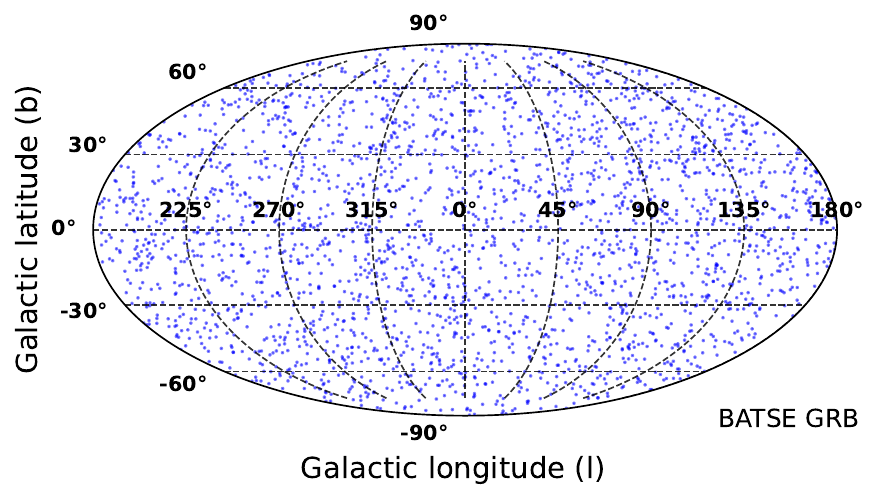}
\centering \includegraphics[width=12cm]{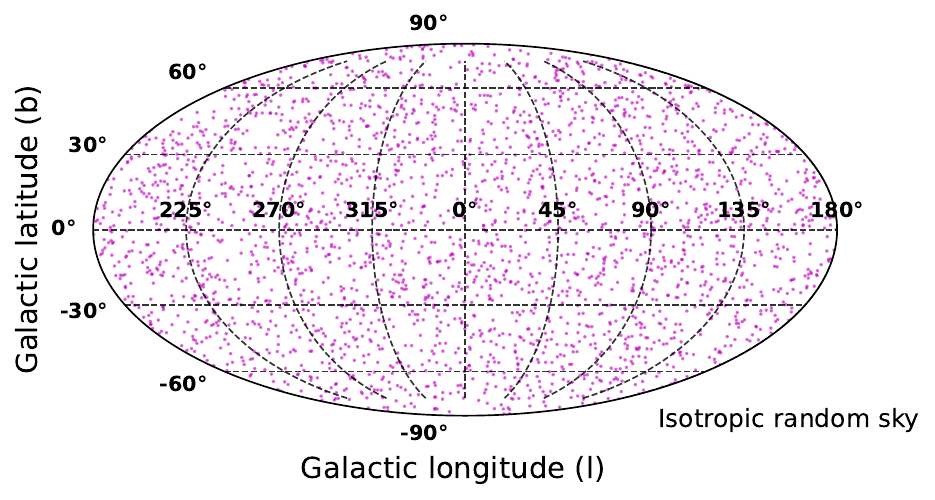} 
\centering \includegraphics[width=12cm]{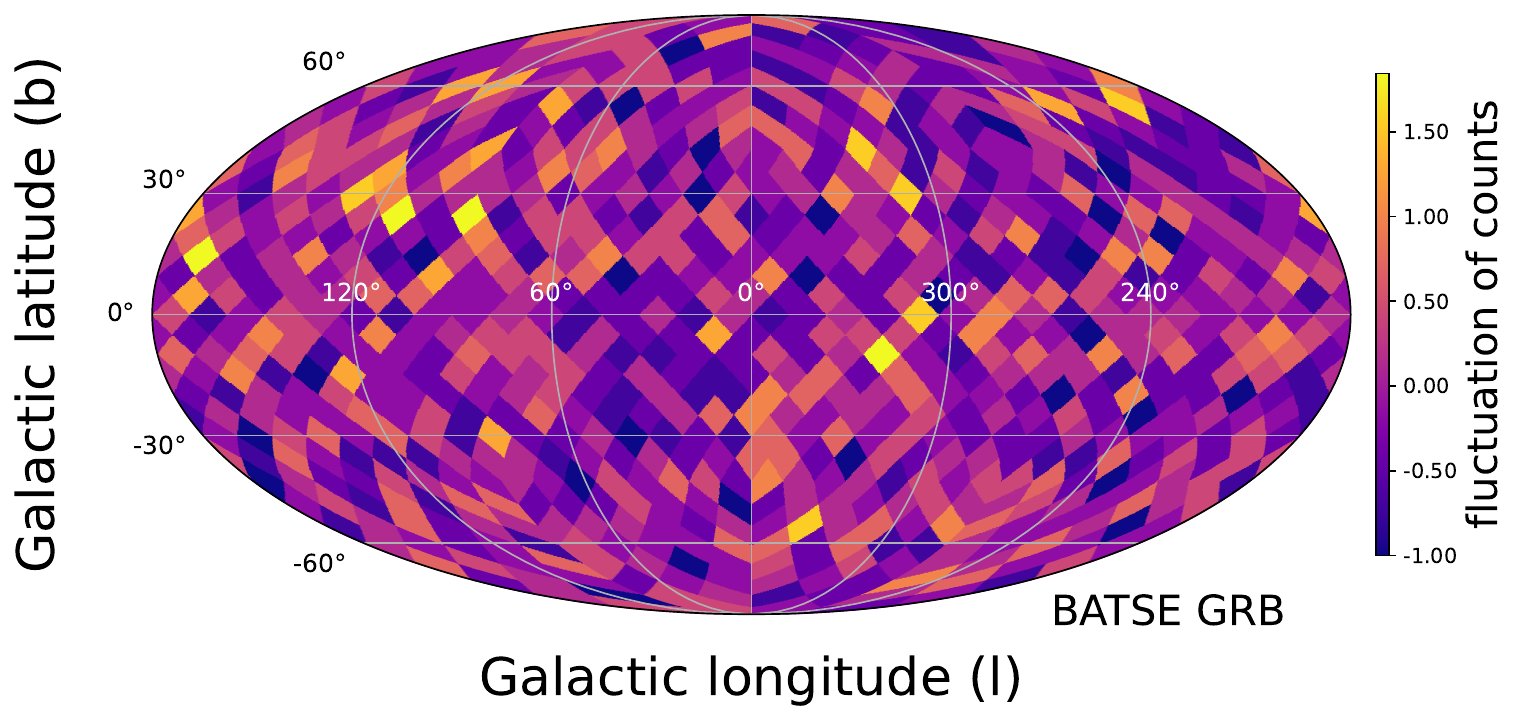} 
\caption{This shows the angular distribution of 2702 GRBs from the BATSE catalog (top panel) and a uniform random distribution of 2702 simulated points (middle panel), shown in Mollweide projection and Galactic coordinates. Bottom panel shows HEALPix map of fluctuation of GRBs from the BATSE catalog using \(\mathrm{N_{side}=8}\).}
\label{fig1}
\end{figure*}

\begin{figure*}[htbp!]
\centering \includegraphics[width=12cm]{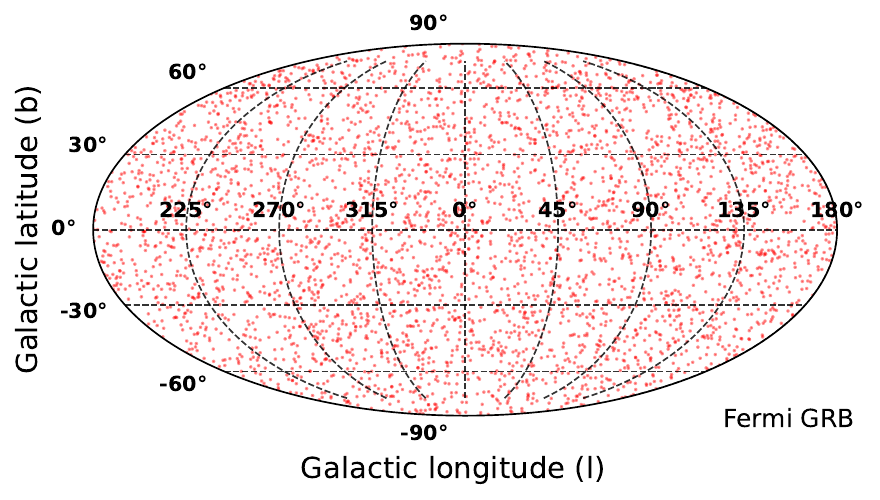}
\centering \includegraphics[width=12cm]{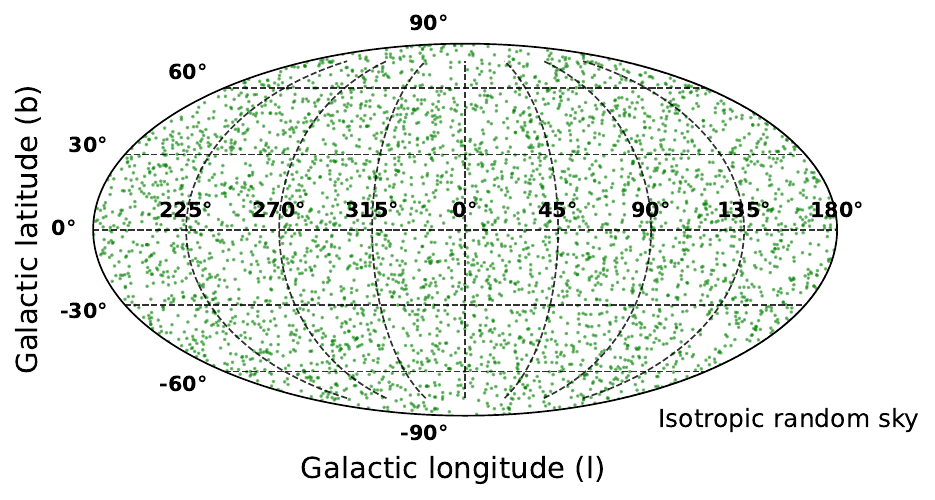} 
\centering \includegraphics[width=12cm]{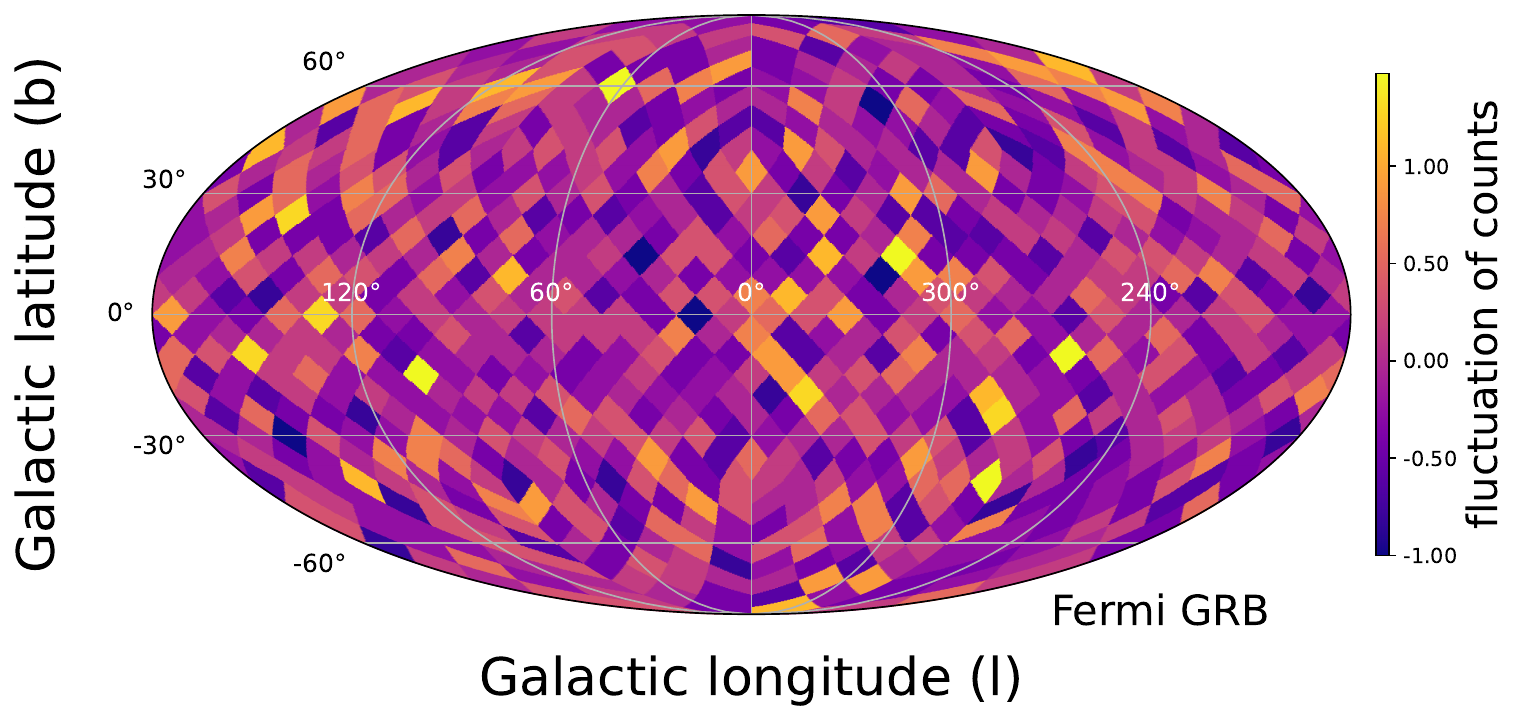} 
\caption{This shows the angular distribution of 4032 GRBs from the Fermi GBM catalog (top panel) and a uniform random distribution of 4032 simulated points (middle panel), shown in Mollweide projection and Galactic coordinates. Bottom panel shows HEALPix map of fluctuation of GRBs from the Fermi GBM catalog using \(\mathrm{N_{side}=8}\).}
\label{fig:2}
\end{figure*}

\section{Data and Method of Analysis}           
\label{sec:data_method}

\subsection{GRB Catalogs}

This study utilizes two of the most comprehensive GRB datasets available: the BATSE Gamma-Ray Burst Catalog and the Fermi GBM Burst Catalog. All data were accessed on June 24, 2025, from publicly available sources.

\subsubsection{BATSE GRB catalog}

Our first GRB dataset is from the \textit{Compton Gamma Ray Observatory (CGRO)/BATSE Gamma-Ray Burst Catalog}\footnote{\url{https://heasarc.gsfc.nasa.gov/W3Browse/cgro/batsegrb.html}} \citep{paciesas99, goldstein13}, which compiles GRBs detected by the Burst and Transient Source Experiment (BATSE) onboard CGRO between 1991 and 2000. This includes events listed in the BATSE 4B Catalog \cite{paciesas99}, as well as additional bursts detected after its publication. The BATSE 3B Catalog \cite{meegan96} is also referenced as an earlier version of the dataset. A total of 2702 GRBs are included, with detection dates ranging from April 21, 1991, to May 26, 2000 (Figure~\ref{fig1}).

\subsubsection{Fermi GBM GRB catalog}

Our second GRB dataset is the \textit{Fermi GBM Burst Catalog}\footnote{\url{https://heasarc.gsfc.nasa.gov/W3Browse/fermi/fermigbrst.html}} \citep{goldstein12, paciesas12, vonkienlin14, gruber14, bhat16, vonkienlin20}, containing GRBs detected by the Gamma-ray Burst Monitor (GBM) onboard the Fermi Gamma-ray Space Telescope, operational since 2008. This continuously updated catalog is among the most complete and homogeneous GRB datasets to date. Our analysis includes 4032 GRBs, with detection dates spanning from July 14, 2008, to June 21, 2025 (Figure~\ref{fig:2}).

\subsection{Method of Analysis}

\subsubsection{Sky pixelation and GRB distribution}

To analyze the angular distribution of GRBs on the celestial sphere, the equatorial coordinates (RA, Dec) were first converted to Galactic coordinates $(l, b)$, where $l \in [0^\circ, 360^\circ]$ and $b \in [-90^\circ, +90^\circ]$.

We employed the Hierarchical Equal Area isoLatitude Pixelization (HEALPix) scheme \cite{gorski05} to divide the sky into equal-area pixels. For a given resolution parameter $N_{\text{side}}$, the sky is partitioned into $N_{\text{pix}} = 12 \times N_{\text{side}}^2$ pixels, each covering an area of $\frac{41253}{N_{\text{pix}}} \text{ deg}^2$. HEALPix ensures each pixel subtends the same solid angle at the center of the celestial sphere.

Using the \texttt{ang2pix} function, we obtained the number count of GRBs in each pixel and then calculated the fluctuations from the mean, resulting in a discretized distribution function $f(\theta, \phi)$, where $\theta \in [0^\circ, 180^\circ]$ and \(\phi \in [0^\circ, 360^\circ]$ represent spherical polar coordinates. The fluctuations  $f(\theta, \phi)$ in each pixel is defined as $\frac{n(\theta, \phi)-\bar{n}}{\bar{n}}$ where $n(\theta, \phi)$ and $\bar{n}$ are number count in $(\theta, \phi)$ direction and mean count respectively.

\subsubsection{Dipole estimation via spherical harmonics}
\label{sec:dipole_estimation}

To test for isotropy, we decomposed the GRB distribution function $f(\theta, \phi)$ into spherical harmonics $Y_{\ell m}(\theta, \phi)$, retaining terms up to the dipole level:

\[
f(\theta, \phi) = \sum_{\ell=0}^{1} \sum_{m=-\ell}^{\ell} a_{\ell m} \, Y_{\ell m}(\theta, \phi)
\]

Here, the monopole term (\(\ell=0\)) reflects uniform distribution, while the dipole term (\(\ell=1\)) captures any directional dependence. The coefficients \(a_{\ell m}\) were calculated using:

\[
a_{\ell m} = \int_{0}^{2\pi} \int_{0}^{\pi} f(\theta, \phi) \, Y_{\ell m}^*(\theta, \phi) \, \sin\theta \, d\theta \, d\phi
\]

The dipole amplitude is given by,

\[
A_{\text{dipole}} = \sqrt{a_{1-1}^2 + a_{10}^2 + a_{11}^2}
\]

We employed the dipole estimator proposed by Secrest et al. \cite{secrest21}, implemented via the \texttt{fit\_dipole} routine from the \texttt{healpy} Python package \cite{zonca19}, which performs a linear least-squares fit of a monopole and dipole model to the HEALPix map.

\subsubsection{Monte Carlo simulations and hypothesis testing}

To evaluate the statistical significance of the observed dipole amplitude, we generated 500 Monte Carlo realizations of isotropically distributed skies, each containing the same number of GRBs as the respective real datasets (2702 for BATSE, 4032 for Fermi GBM). For each simulated sky, we calculated the dipole amplitude using the same methodology described above.

These simulated amplitudes were then used to construct a probability density function (PDF) for the dipole amplitude under the assumption of isotropy. The PDFs were smoothed using Kernel Density Estimation (KDE).

We then compared the dipole amplitude from the observed GRB data with the corresponding PDF to assess statistical significance. This constitutes a one-tailed hypothesis test, with the null hypothesis \(H_0\) stating that the GRB sky is isotropic and any observed dipole arises from random fluctuations.

The significance level was set at $\alpha = 0.05$ (corresponding to a 95\% confidence level). The $p$-value was computed as $p = P(X \geq x_{\text{obs}}) = \int_{x_{\text{obs}}}^{\infty} f(x) \, dx$. If \(p < \alpha\), we reject the null hypothesis, indicating statistically significant anisotropy. Conversely, if $p \geq \alpha$, we conclude that the observed dipole amplitude is consistent with statistical fluctuations in an isotropic sky.

\subsubsection{Validation with simulated dipole skies}
\label{sec:dipole_validation}

To assess the reliability of our dipole analysis framework, we perform a controlled validation by generating simulated GRB skies with injected dipole anisotropy. These synthetic skies are constructed to contain the same number of points as the real BATSE catalog (2702 GRBs), but with a statistically imposed directional dependence. We then apply our full analysis pipeline to these mock datasets to determine whether the injected dipole signals can be recovered with statistical significance.

To simulate a GRB sky with dipole modulation, we use the following dipole-modified probability density function,
\begin{equation}
    P(\theta, \phi) = N \left(1 + a \cos \gamma \right),
    \label{eq:dipole_pdf}
\end{equation}
where $a$ is the dipole strength parameter, $N$ is a normalization constant ensuring the total probability integrates to one over the sphere, and $\gamma$ is the angle between the dipole direction $(\theta_{\text{dipole}}, \phi_{\text{dipole}})$ and the point $(\theta, \phi)$ on the sky. The angle $\gamma$ is computed using the spherical law of cosines:
\begin{equation}
\cos \gamma = \cos \theta \cos \theta_{\text{dipole}} + \sin \theta \sin \theta_{\text{dipole}} \cos(\phi - \phi_{\text{dipole}}).
\end{equation}

To ensure that the probability distribution remains valid, i.e., $P(\theta, \phi) > 0$, we restrict the dipole amplitude to $|a| < 1$.

We generate random \((\theta, \phi)\) pairs uniformly over the sphere and evaluate \(P(\theta, \phi)\) for each point. A $(\theta, \phi)$ pair is accepted if a uniformly drawn random number $u \in [0, P_{\text{max}}]$ satisfies $u < P(\theta, \phi)$. Otherwise, we reject the $(\theta, \phi)$ pair. This rejection sampling is repeated until 2702 accepted points are obtained, resulting in a simulated sky with a specified dipole anisotropy.

For our validation, we set the dipole direction to $(\theta_{\text{dipole}}, \phi_{\text{dipole}}) = (40^\circ, 140^\circ)$ and generate four dipole-injected skies with amplitudes $a = 0.09, 0.10, 0.12,$ and $0.14$ in \autoref{eq:dipole_pdf}. For each realization, we compute the dipole amplitude using the same methodology as described in Section~\ref{sec:dipole_estimation}, and compare the results with the null distribution obtained from isotropic Monte Carlo simulations.

\begin{figure*}[htbp!]
\centering
\includegraphics[width=12cm]{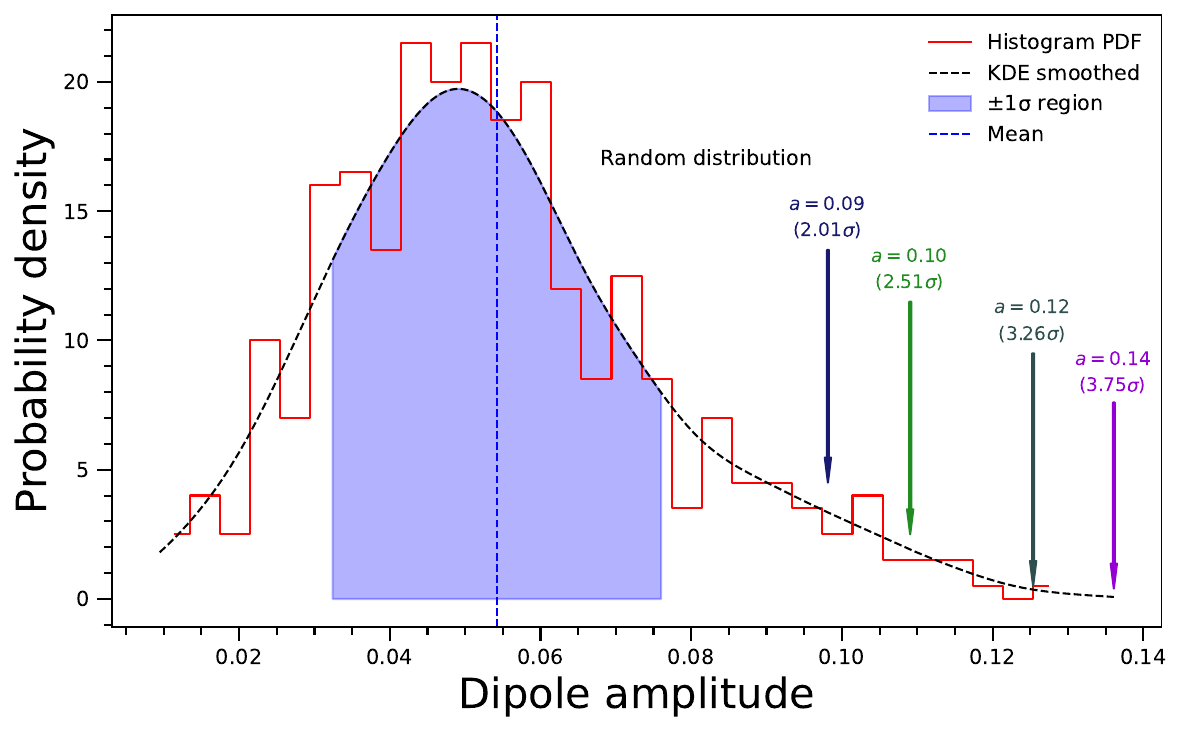}
\caption{Comparison of dipole amplitudes from four simulated dipole skies ($a = 0.09, 0.10, 0.12, 0.14$) with the estimated PDF obtained from $500$ Monte Carlo simulations of isotropic skies (2702 GRBs each). All injected dipole amplitudes lie beyond the 1$\sigma$ region of the null distribution. The HEALPix resolution parameter \(\mathrm{N_{side}}=8\) has been used for this analysis.  }
\label{fig:5}
\end{figure*}

Figure~\ref{fig:5} shows the dipole amplitudes of the four simulated skies superimposed on the PDF constructed from 500 isotropic sky realizations. Table~\ref{tab:dipole_significance} summarizes the corresponding \(p\)-values and their statistical significance in terms of standard deviations ($\sigma$) from the mean of the null distribution.

\begin{table}[ht]
\centering
\begin{tabular}{cccc}
\hline
Sl No. & \(\mathbf{a}\) & $p$-value & Significance (in $\sigma$) \\
\hline
1 & 0.09 & 0.0479 & 2.01 \\
2 & 0.10 & 0.0193 & 2.51 \\
3 & 0.12 & 0.0024 & 3.26 \\
4 & 0.14 & 0.0003 & 3.75 \\
\hline
\end{tabular}
\caption{Statistical significance of the recovered dipole amplitudes for simulated skies with various injected dipole strengths \(a\). All \(p\)-values fall below \(\alpha = 0.05\).}
\label{tab:dipole_significance}
\end{table}

The results confirm that our method reliably detects dipole anisotropy across a range of amplitudes. For all four cases, the observed dipole amplitudes lie well outside the region expected from random fluctuations under the null hypothesis. Even the lowest injected amplitude of \(a = 0.09\) yields a significance level above \(2\sigma\), while \(a = 0.14\) is detected at nearly \(3.8\sigma\). This demonstrates that our framework is sensitive enough to identify dipole signals of modest strength and can robustly distinguish them from statistical noise.

\begin{figure*}[htbp!]
\centering
\includegraphics[width=12cm]{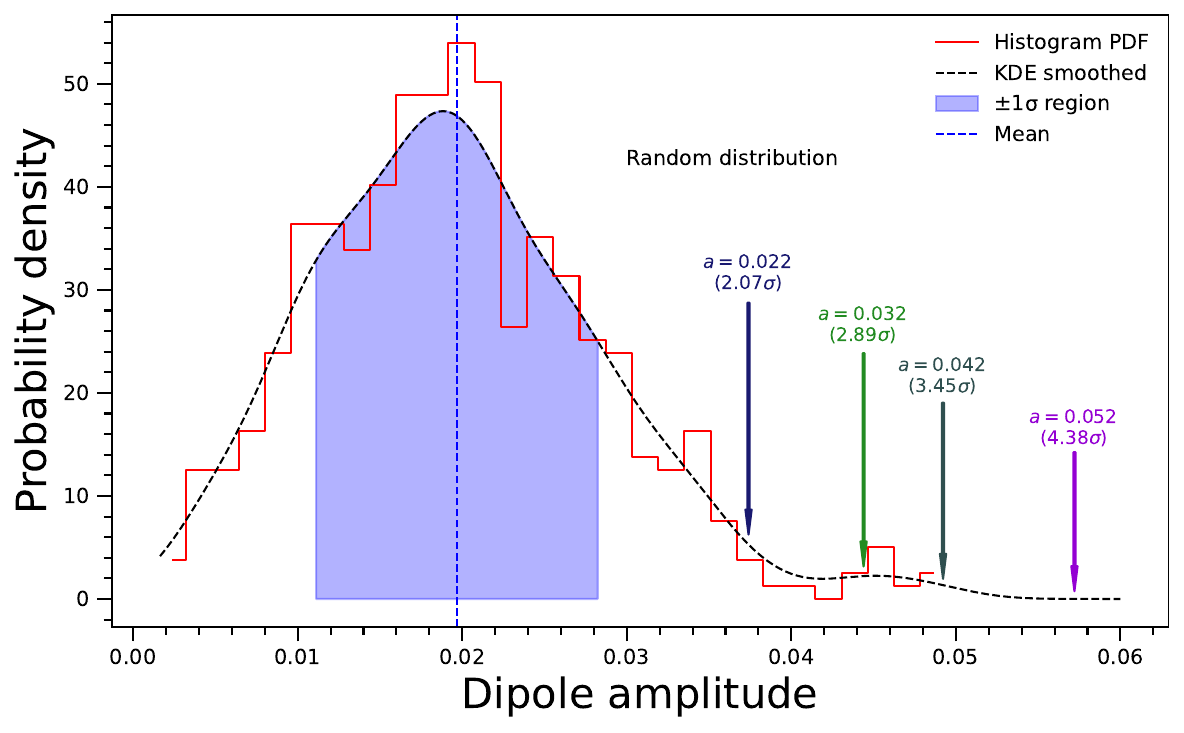}
\caption{Same as \autoref{fig:5}, but for simulated isotropic and dipole skies each containing $20{,}000$ points. Four dipole skies with amplitudes $a = 0.022, 0.032, 0.042,$ and $0.052$ were generated, maintaining the same dipole direction used in \autoref{fig:5}.}
\label{fig:6}
\end{figure*}

We further examined the detection efficiency of our method by extending the analysis to larger simulated datasets. Specifically, we generated both isotropic and dipole-modulated skies, each containing $20000$ points, and compared their dipole amplitudes to assess how sample size influences sensitivity. Following the same procedure as before, we constructed the dipole amplitude distribution from 500 isotropic realizations and compared it against four dipole-injected skies with amplitudes $a = 0.022,\, 0.032,\, 0.042$, and $0.052$, oriented in the same direction as in the previous test. The corresponding probability distributions and significance levels are shown in \autoref{fig:6} and summarized in \autoref{tab:dipole_significance2}.

\begin{table}[ht]
\centering
\begin{tabular}{cccc}
\hline
Sl no. & \(\mathbf{a}\) & $p$-value & Significance (in $\sigma$) \\
\hline
1 & 0.022 & 0.0312 & 2.07 \\
2 & 0.032 & 0.0125 & 2.89 \\
3 & 0.042 & 0.0030 & 3.45 \\
4 & 0.052 & 2.387 $\times 10^{-6}$ & 4.38 \\
\hline
\end{tabular}
\caption{Same as \autoref{tab:dipole_significance}, but for simulations containing $20000$ points.}
\label{tab:dipole_significance2}
\end{table}

The results clearly demonstrate that increasing the number of data points enhances the sensitivity of the analysis. In this higher-resolution test, even a weak dipole signal with amplitude $a = 0.022$ lies beyond the $2\sigma$ region of the isotropic distribution, confirming that the method can reliably detect subtle anisotropies when applied to sufficiently large datasets. This underscores that the efficiency of dipole detection depends not only on the signal strength but also on the statistical size of the sample representing the GRB sky. Larger simulated samples would allow the detection of progressively weaker dipole signals at a fixed confidence level. The choice of $20000$ points in our validation tests is intended to illustrate the behavior of the method in a high-statistics regime, beyond the sizes of the current BATSE and Fermi GBM catalogs. Using a smaller number of events (e.g., $10000$) would require a correspondingly larger dipole amplitude to achieve the same level of statistical significance, while larger samples (e.g., $50000$ points) would enable sensitivity to even weaker anisotropies. Varying the number of simulated points does not alter the qualitative performance of the method, but instead produces the expected scaling of detection thresholds with sample size.

\section{Results}
\label{sec:results}

\subsection{Dipole amplitude analysis of BATSE dataset}

To assess the statistical isotropy of the BATSE GRB sky, we calculated the dipole amplitude of the observed distribution and compared it with a reference distribution constructed from 500 Monte Carlo realizations of isotropic skies, each containing 2702 GRBs equal to the number in the actual BATSE dataset.

\begin{figure*}[htbp!]
\centering
\includegraphics[width=0.9\textwidth]{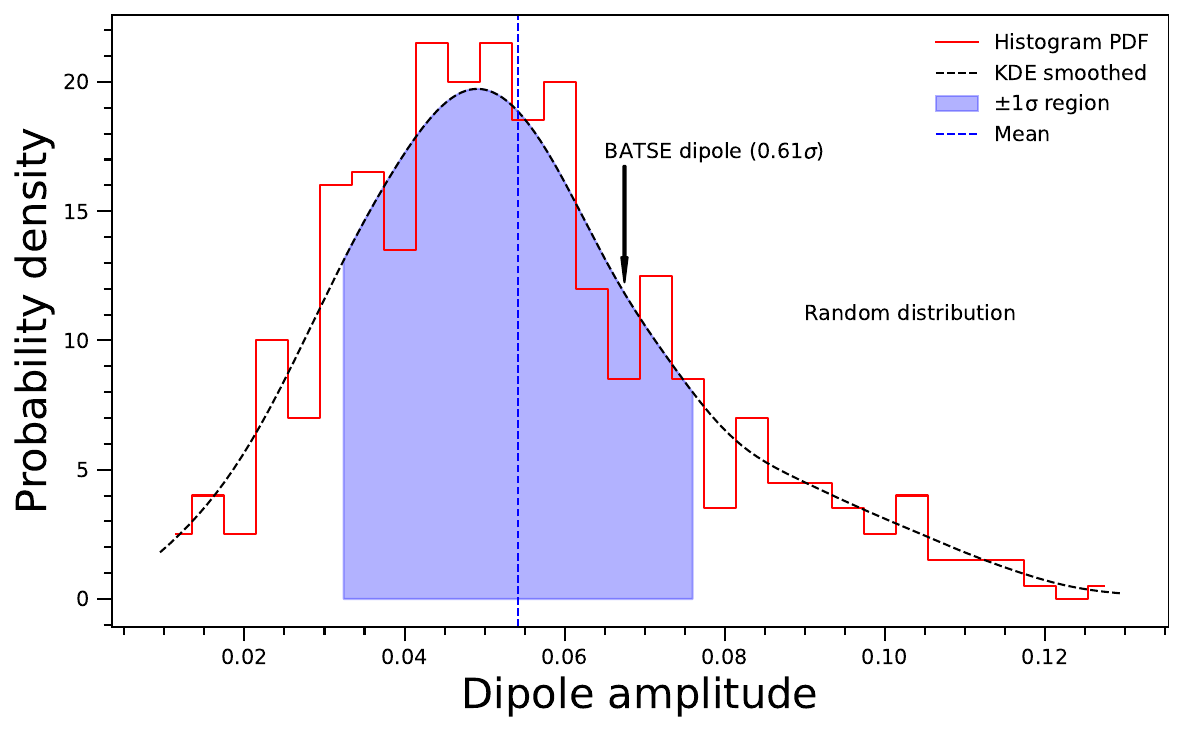}
\caption{ This shows the comparison of the observed dipole amplitude for the BATSE dataset with the PDF obtained from Monte Carlo simulations of 500 isotropic skies containing 2702 points each. The red step histogram shows the binned PDF, while the dashed black line represents the KDE-smoothed curve. The shaded region marks the $\pm1\sigma$ interval around the mean. The vertical dashed blue line represents the mean of the distribution and the downward black arrow indicates the observed BATSE dipole amplitude, which lies within the $1\sigma$ range, suggesting no significant deviation from isotropy. The analysis was performed using a HEALPix resolution parameter of $\mathrm{N_{side}} = 8$.}
\label{fig:batse_pdf}
\end{figure*}

Figure~\ref{fig:batse_pdf} presents the resulting PDF of dipole amplitudes derived from the simulated isotropic skies. The red step histogram represents the binned distribution of dipole amplitudes across these simulations, while the dashed black curve shows the kernel density estimation (KDE), which provides a smoothed representation of the underlying PDF. The shaded blue region indicates the \(1\sigma\) interval around the mean of the distribution.

The observed dipole amplitude from the BATSE sky is indicated by the vertical blue dashed line and a downward-pointing arrow labeled ``BATSE dipole (0.61$\sigma$)''. As evident from the figure, the observed value lies well within the \(1\sigma\) range of the null distribution. This placement suggests that the dipole amplitude measured from the BATSE dataset is consistent with expectations from random statistical fluctuations in an isotropic sky.

The associated $p$-value, calculated as the probability of obtaining a dipole amplitude equal to or greater than the observed value under the null hypothesis, exceeds the significance threshold of $\alpha = 0.05$. Therefore, we do not reject the null hypothesis of isotropy. In other words, the analysis reveals no statistically significant evidence of large-scale dipole anisotropy in the BATSE GRB distribution.

\subsection{Dipole amplitude analysis of Fermi GBM dataset}

We perform a similar dipole amplitude analysis for the Fermi GBM dataset, which includes 4032 GRBs detected between 2008 and 2025. As with the BATSE analysis, we compare the dipole amplitude calculated from the observed sky distribution with a null distribution constructed from 500 isotropic Monte Carlo realizations, each containing the same number of GRBs as the Fermi sample.

\begin{figure*}[htbp!]
\centering
\includegraphics[width=0.9\textwidth]{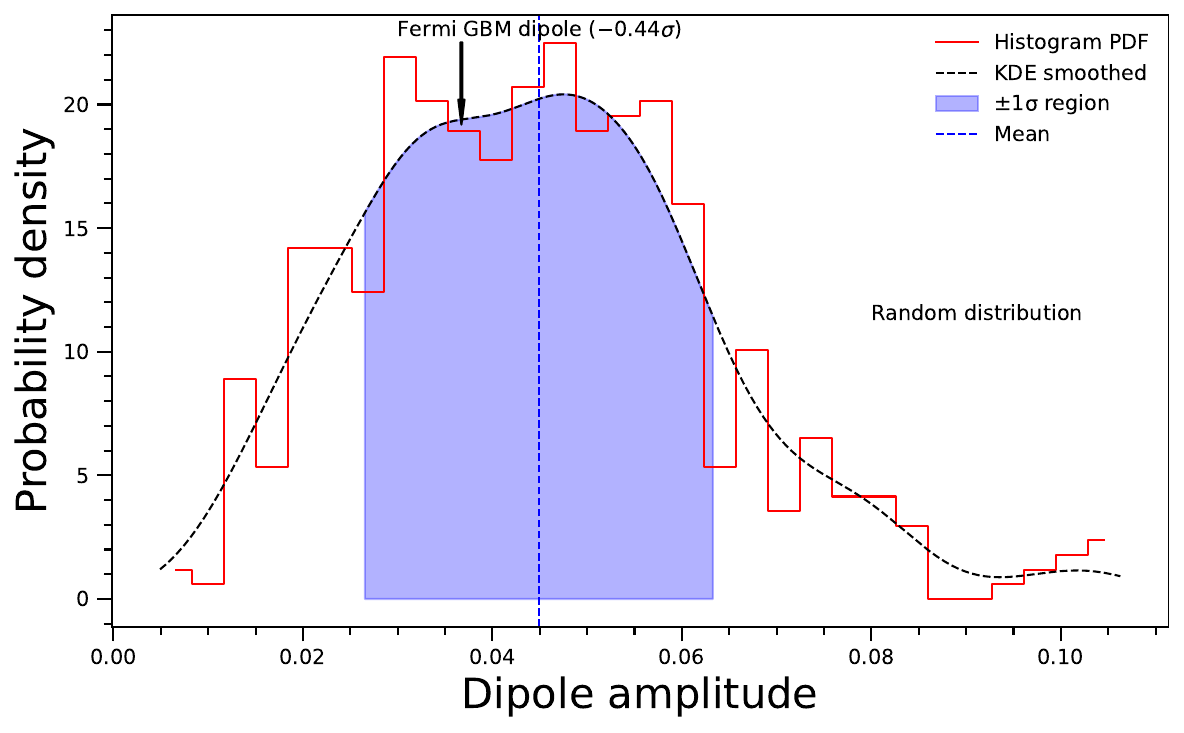}
\caption{Same as Figure~\ref{fig:batse_pdf} but for Fermi GBM dataset.}
\label{fig:fermi_pdf}
\end{figure*}

Figure~\ref{fig:fermi_pdf} illustrates the distribution of dipole amplitudes derived from the isotropic simulations. The red step histogram represents the binned PDF, while the black dashed curve shows the kernel density estimate (KDE), providing a smooth approximation of the underlying distribution. The shaded blue region marks the \( \pm1\sigma \) interval around the mean of the null distribution.

The dipole amplitude computed from the Fermi GBM data is shown by the downward-pointing black arrow, labeled ``FERMI GBM dipole ($-0.44 \sigma$)''. Notably, the observed value lies well within the 1$\sigma$ range of the null distribution, and in fact falls slightly below the mean value derived from the simulations.

This result is consistent with expectations from an isotropic sky and suggests no statistically significant anisotropy in the angular distribution of Fermi-detected GRBs. The associated $p$-value is well above the significance threshold of $\alpha = 0.05$, leading us to retain the null hypothesis. In other words, the dipole amplitude observed in the Fermi dataset is entirely consistent with random statistical fluctuations expected under isotropy.

\subsection{Quadrupole amplitude analysis of the BATSE dataset}

To further probe possible anisotropic features in the angular distribution of GRBs, we extended our analysis beyond the dipole term to include the quadrupole component. The quadrupole moment, sensitive to planar or axial anisotropies in the sky distribution, was computed using the \texttt{anafast} routine of \texttt{Healpy}, which estimates the angular power spectrum coefficients \(C_{\ell}\) from HEALPix maps. In this context, the quadrupole power \(C_{2}\) represents the variance of the sky distribution at multipole order \(\ell = 2\), and its square root provides a convenient measure of the quadrupole amplitude.

\begin{figure*}[htbp!]
\centering
\includegraphics[width=12cm]{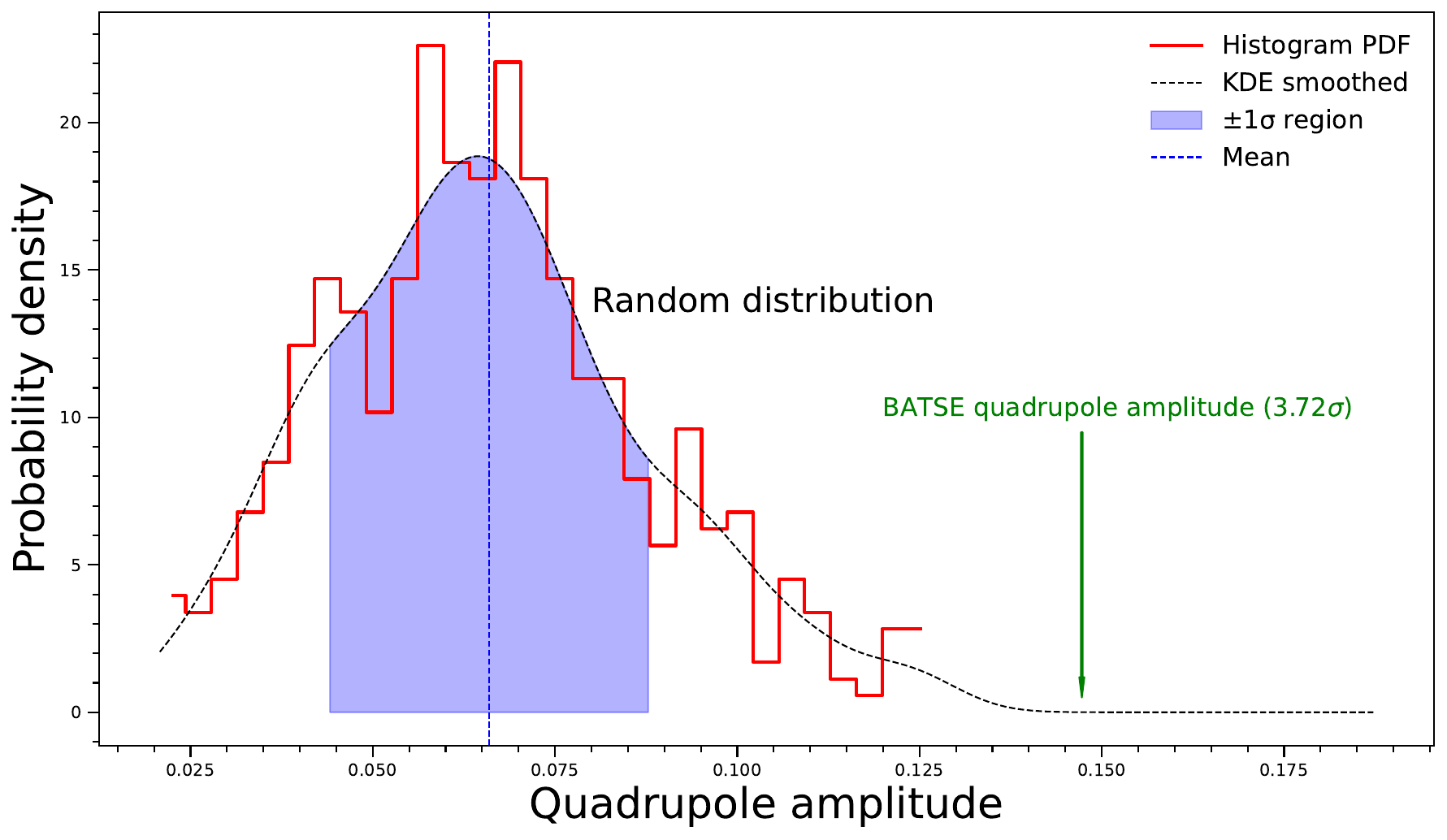}
\caption{Comparison of the observed quadrupole amplitude for the BATSE dataset (vertical downward arrow) with the PDF constructed from Monte Carlo simulations of 500 isotropic skies, each containing 2702 points. The observed quadrupole amplitude lies approximately $3.72\sigma$ above the mean of the isotropic distribution, corresponding to a $p$-value of $4.306 \times 10^{-6}$. The analysis was performed using $\mathrm{N_{side}} = 8$.}
\label{fig:7}
\end{figure*}

Following the same Monte Carlo approach described previously, we generated 500 isotropic skies containing the same number of points as in the BATSE catalog (2702 GRBs). The quadrupole amplitudes of these simulated skies were used to construct a PDF, against which the quadrupole amplitude of the real BATSE sky was compared. The resulting comparison is shown in \autoref{fig:7}.

The results indicate that the observed quadrupole amplitude in the BATSE GRB distribution is significantly higher than expected from random isotropic realizations. The deviation, corresponding to a statistical significance of approximately $3.7\sigma$, suggests the presence of a weak but non-negligible large-scale anisotropy. While this finding is intriguing, it should be interpreted cautiously, as residual instrumental effects, exposure non-uniformities, or catalog selection biases could potentially contribute to the observed excess. Nevertheless, the result warrants further investigation using independent datasets and higher-resolution analyses to assess the persistence and physical origin of this quadrupole feature.

\subsection{Quadrupole amplitude analysis of the FERMI dataset}

To complement the quadrupole analysis performed on the BATSE catalog, we extended the investigation to the Fermi GBM GRB dataset, which provides a larger and more recent sample of bursts with improved all-sky coverage. Using the same methodology, we employed the \texttt{anafast} routine of \texttt{Healpy} to compute the quadrupole power, \(C_{2}\), from the HEALPix map representation of the Fermi GRB sky. The square root of \(C_{2}\) was taken as a measure of the quadrupole amplitude, providing a direct assessment of any large-scale planar or axial anisotropies present in the GRB distribution.

\begin{figure*}[htbp!]
\centering
\includegraphics[width=12cm]{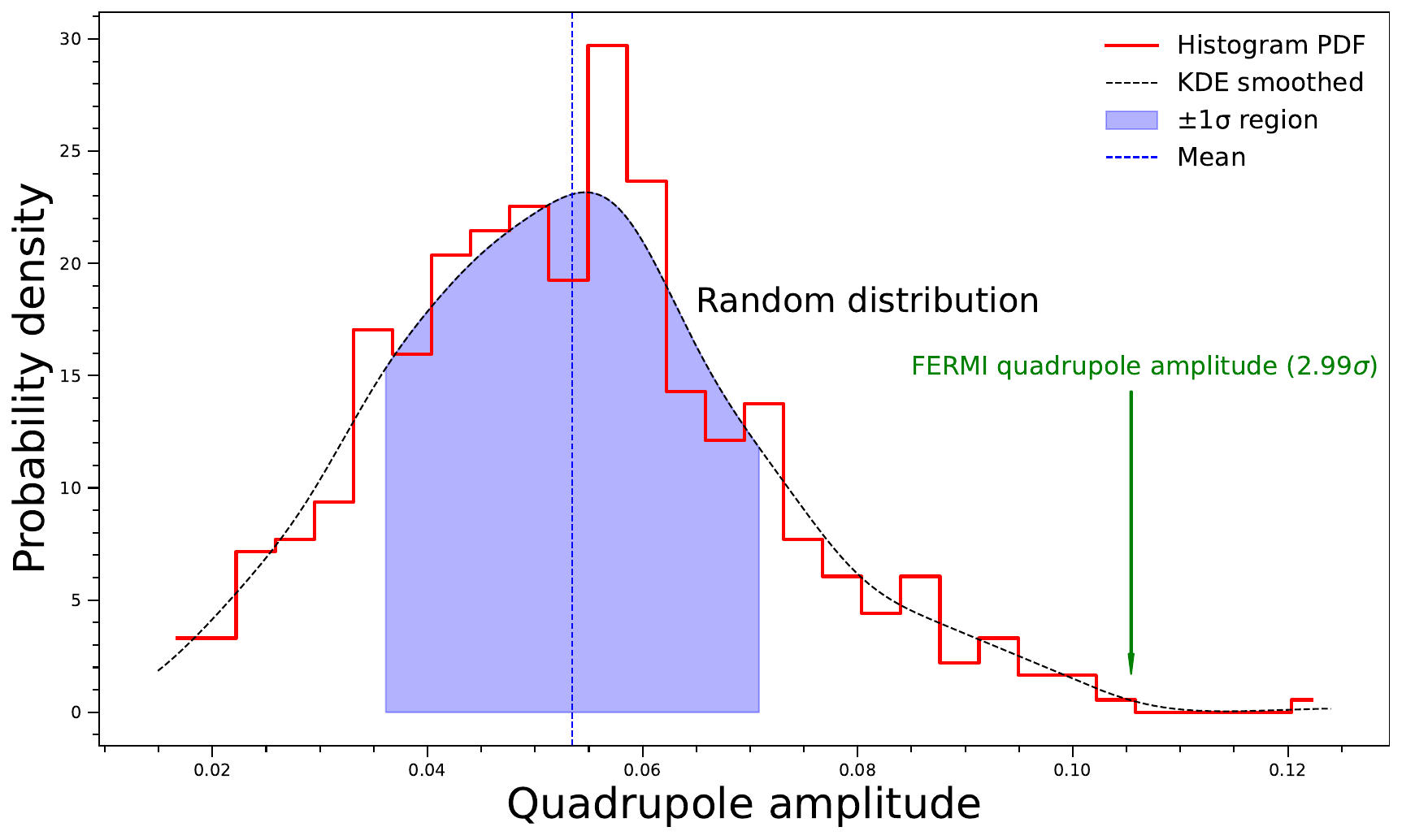}
\caption{Same as Figure~\ref{fig:7} but for Fermi GBM dataset.}
\label{fig:8}
\end{figure*}

A set of 500 Monte Carlo realizations of isotropic skies, each containing 4032 points (matching the size of the Fermi sample), was generated to establish the null distribution of quadrupole amplitudes. The resulting PDF of the simulated quadrupole amplitudes was compared with the measured value from the real Fermi data, as shown in \autoref{fig:8}.

The Fermi GBM results reveal a quadrupole amplitude moderately higher than the expectation from purely isotropic skies, with a significance level of about $3\sigma$. This deviation, though less pronounced than that found in the BATSE data, still suggests the presence of a weak large-scale anisotropy in the GRB distribution. The consistency of a non-zero quadrupole signal across two independent GRB datasets, spanning different observational periods and instrumental configurations, may hint at a genuine cosmological origin rather than an observational artifact. However, given that both instruments have different sky exposure patterns and systematic limitations, further analysis particularly with future, uniformly sampled GRB catalogs will be essential to confirm the persistence and nature of this quadrupole signature.

\subsection{Testing the significance of the dipole and quadrupole after applying exposure correction to the BATSE dataset}
\label{sec:batse_exposure_correction}

To assess whether the observed large-scale anisotropies in the BATSE Gamma-Ray Burst (GRB) distribution arise from intrinsic cosmic effects or instrumental biases, we performed a detailed analysis incorporating exposure corrections. The sky exposure of the BATSE detector onboard the Compton Gamma Ray Observatory (CGRO) was not uniform, varying primarily with declination due to the spacecraft’s orbit and observing constraints. Such exposure non-uniformity can artificially imprint large-scale anisotropies particularly in the dipole and quadrupole moments on the observed GRB sky distribution. To quantify this effect, we employed the BATSE 5B exposure function \citep{hakkila03}, which describes the exposure as a function of declination and remains independent of right ascension (RA). Consequently, all subsequent analyses were performed in equatorial coordinates, where this dependence is naturally represented.

\begin{figure*}[htbp!]
\centering
\includegraphics[width=12cm]{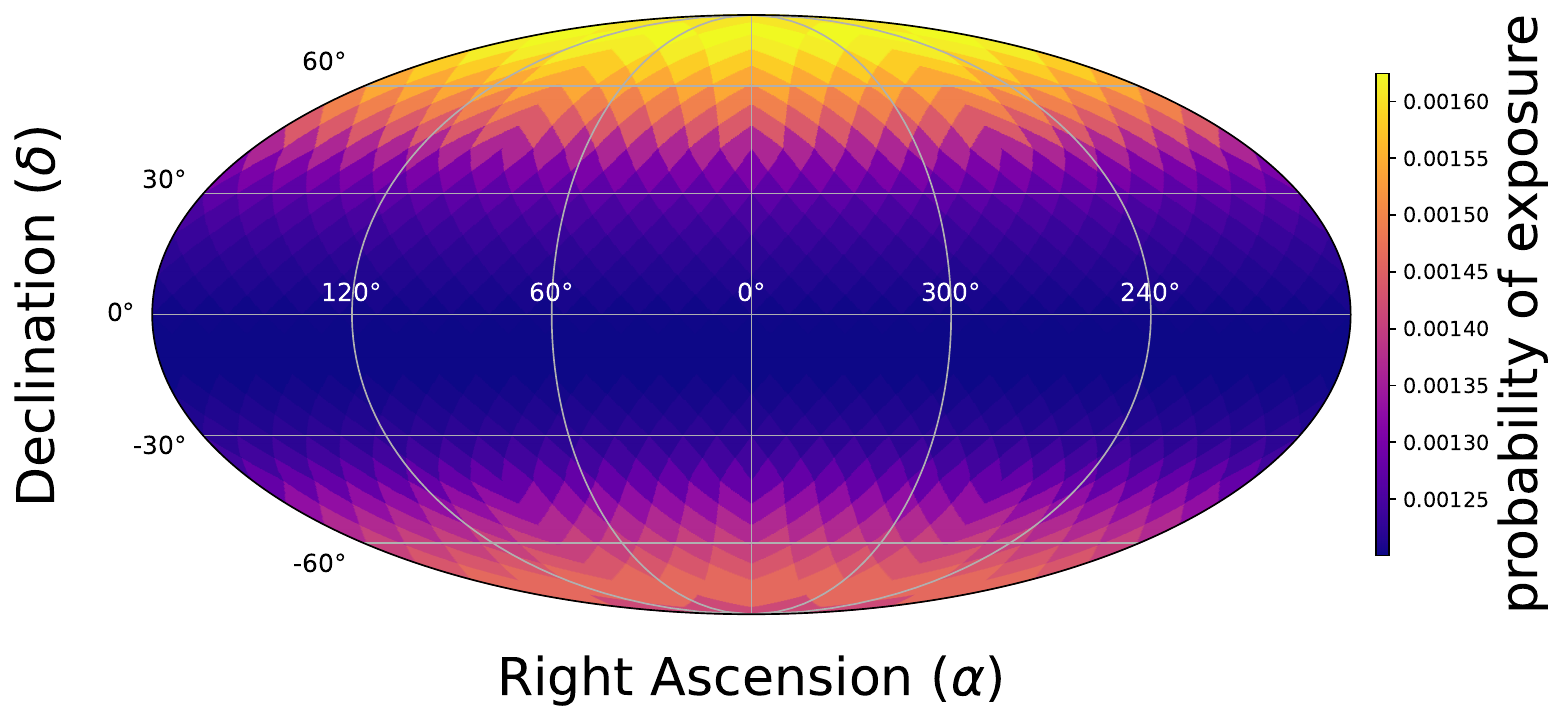}
\caption{HEALPix probability map of BATSE sky exposure in equatorial coordinates using $\mathrm{N_{side}}=8$.}
\label{fig:9}
\end{figure*}

\begin{figure*}[htbp!]
\centering
\includegraphics[width=12cm]{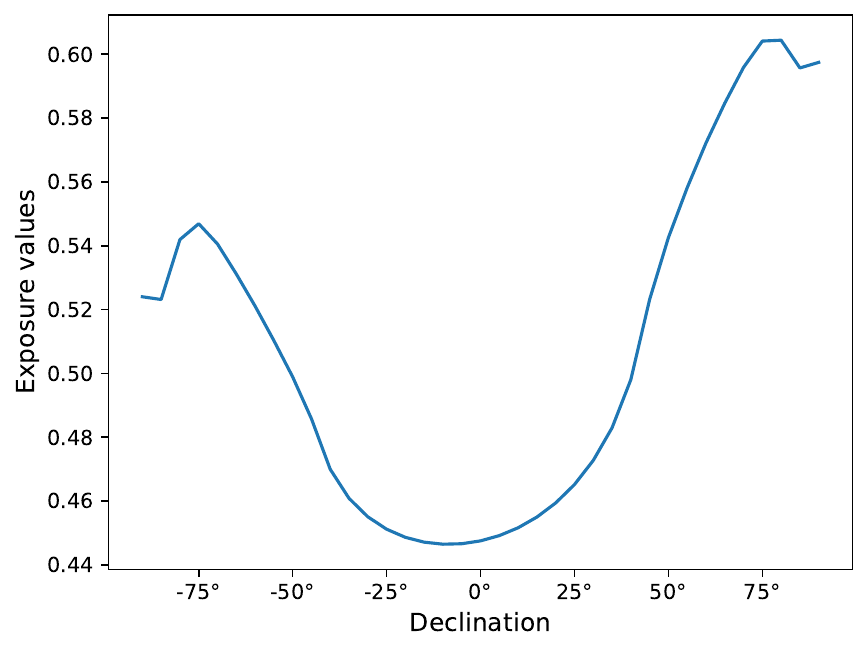}
\caption{BATSE 5B exposure function as a function of declination. The exposure peaks near the poles and decreases toward the equatorial regions, reflecting the instrument’s observing geometry.}
\label{fig:10}
\end{figure*}

We generated mock isotropic skies convolved with the BATSE exposure map to serve as a reference for statistical comparison. The procedure is outlined as follows. The full sky was pixelized using the HEALPix scheme with a resolution parameter $\mathrm{N_{side}} = 8$, resulting in $12 \times \mathrm{N_{side}}^2$ pixels. The declination value for each pixel was determined using $\mathrm{Dec} = 90^\circ - \theta$, where $\theta$ is the polar angle. Exposure values were then assigned to each pixel through linear interpolation of the BATSE 5B exposure data, producing a complete HEALPix map of the exposure function (\autoref{fig:9} and \autoref{fig:10}). These exposure values were normalized to form a probability map, ensuring that the integrated probability across the entire sky equals unity.

\begin{figure*}[htbp!]
\centering
\begin{subfigure}{12cm}
\centering
\includegraphics[width=12cm]{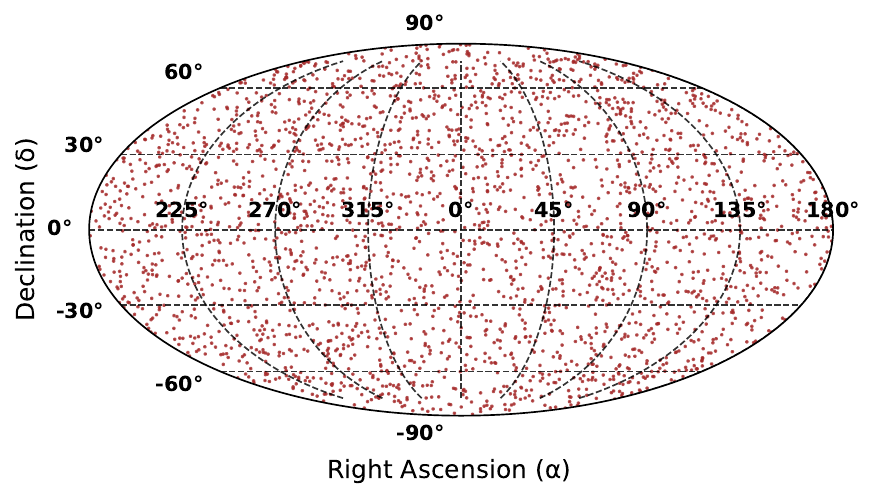}
\subcaption{Distribution of GRB positions in a mock exposure-convolved isotropic sky shown in Mollweide projection (equatorial coordinates).}
\end{subfigure}
\vspace{0.5cm}
\begin{subfigure}{12cm}
\centering
\includegraphics[width=12cm]{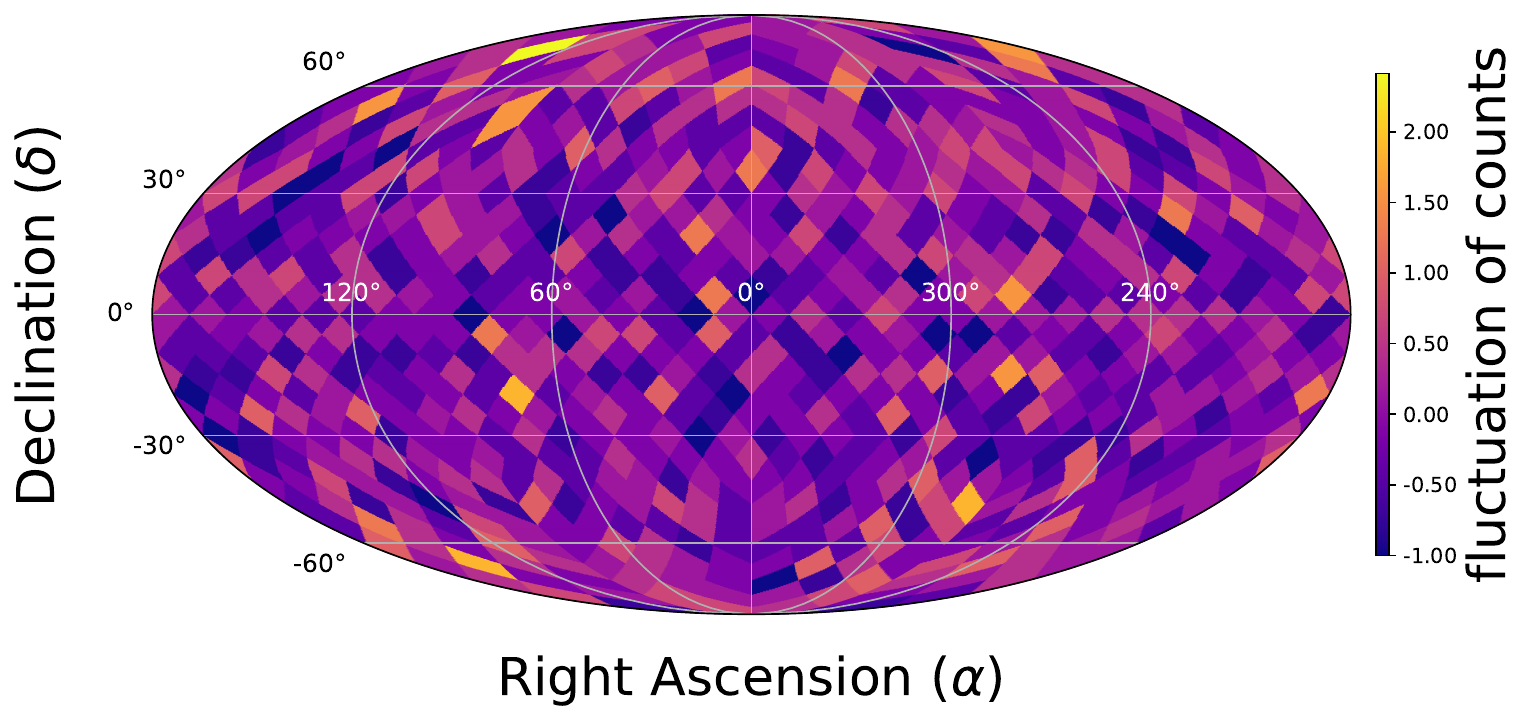}
\subcaption{HEALPix map of GRB count fluctuations in a mock exposure-convolved isotropic sky using $\mathrm{N_{side}}=8$.}
\end{subfigure}
\caption{Simulated exposure-convolved isotropic skies constructed using the BATSE 5B exposure function. These maps illustrate the modulation of sky uniformity induced by the instrument’s varying exposure.}
\label{fig:11}
\end{figure*}

Uniformly distributed $(\theta, \phi)$ points were subsequently generated on the celestial sphere. For each point, a random number $u$ was drawn from a uniform distribution between 0 and the maximum exposure probability. A point was accepted if $u$ was less than the corresponding exposure probability value, thereby ensuring that the generated GRB positions follow the BATSE exposure pattern. This procedure was repeated until 2702 GRBs (matching the size of the observed BATSE catalog) were produced for each of the 500 mock skies. Representative examples of these exposure-convolved isotropic skies are shown in \autoref{fig:11}.

\begin{figure*}[htbp!]
\centering
\includegraphics[width=12cm]{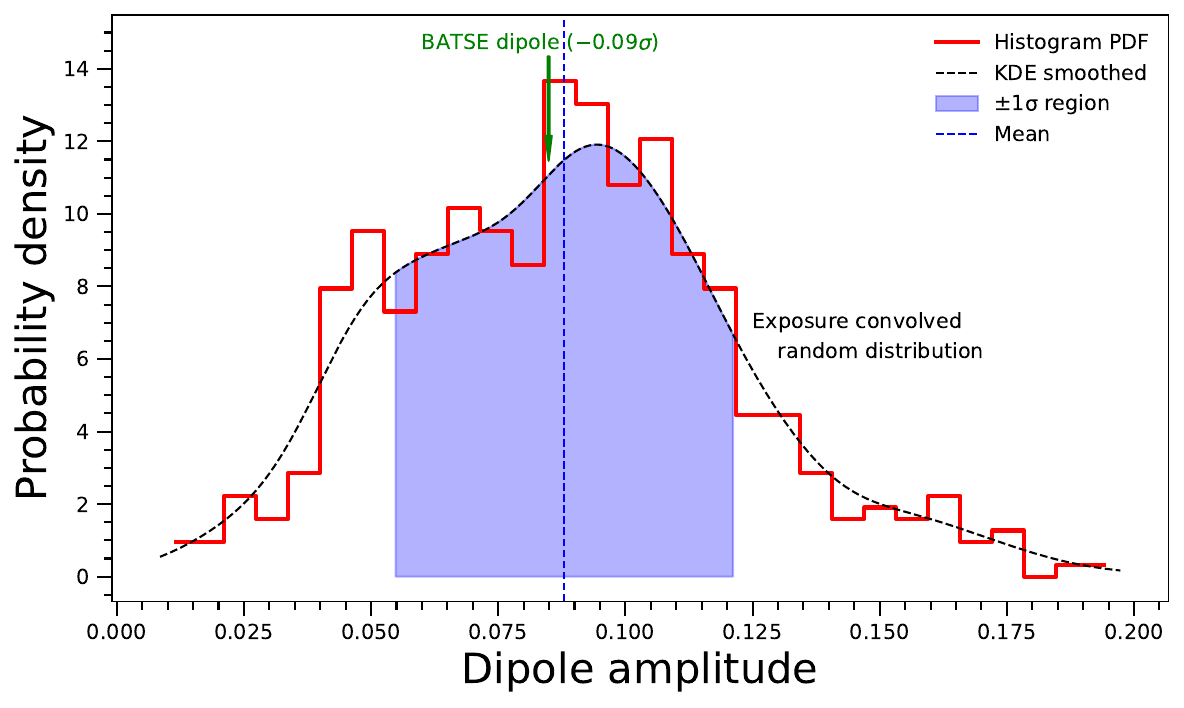}
\caption{This compares the observed dipole amplitude for BATSE data (vertical downward arrow) with the PDF derived from Monte Carlo simulations of 500 exposure-convolved isotropic skies containing 2702 points each. The observed dipole amplitude lies within $0.1\sigma$ of the mean, indicating no statistically significant dipole anisotropy. The HEALPix resolution parameter used is $\mathrm{N_{side}}=8$.}
\label{fig:12}
\end{figure*}

We compared the dipole and quadrupole amplitudes of the real BATSE sky with the corresponding PDFs derived from the 500 exposure-convolved isotropic simulations. The dipole results are presented in \autoref{fig:12}. Once the exposure correction was applied, the apparent dipole signal vanished. The observed dipole amplitude now lies only $-0.09\sigma$ away from the mean of the isotropic distribution. This demonstrates that the previously detected dipole anisotropy in the uncorrected BATSE data was predominantly driven by the non-uniform sky exposure, rather than by an intrinsic cosmological dipole.

\begin{figure*}[htbp!]
\centering
\includegraphics[width=12cm]{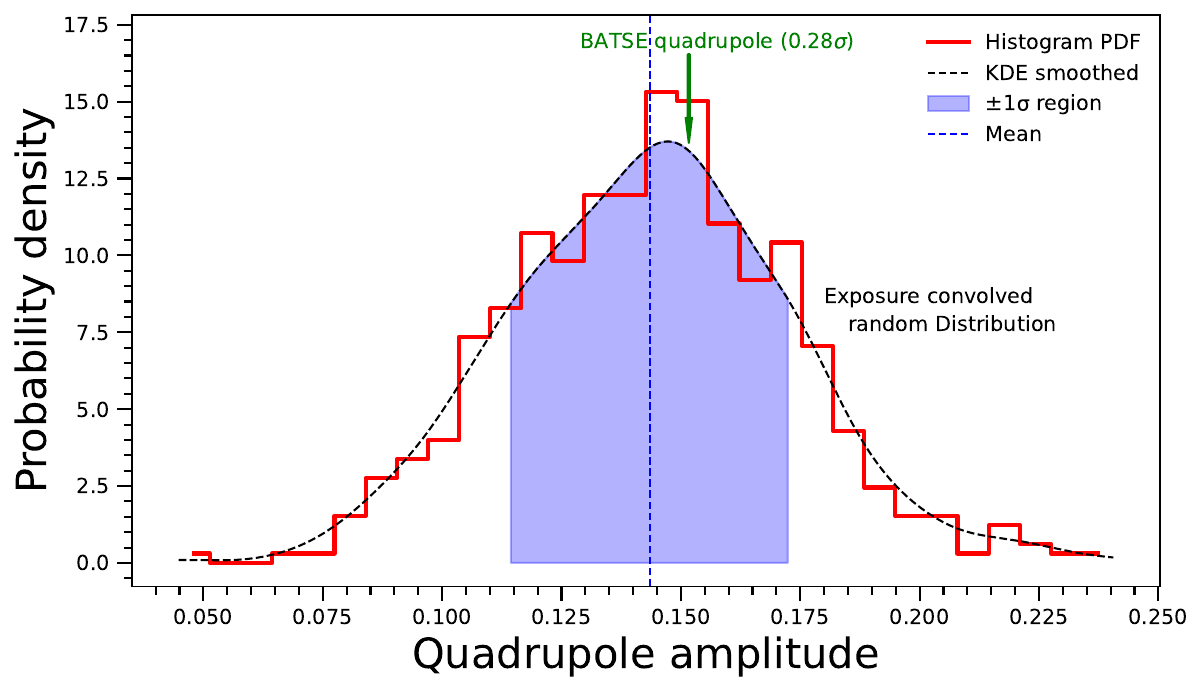}
\caption{Comparison of the observed quadrupole amplitude for BATSE data (vertical downward arrow) with the
estimated PDF obtained from Monte Carlo simulations of 500 exposure-convolved isotropic
skies containing 2702 points each. The observed quadrupole amplitude lies $0.28\sigma$ away from the mean of the isotropic distribution, indicating no significant deviation from isotropy. The HEALPix resolution parameter used is $\mathrm{N_{side}}=8$.}
\label{fig:15}
\end{figure*}

A similar trend is observed for the quadrupole moment. Using the \texttt{anafast} routine of \texttt{Healpy}, we computed the quadrupole power ($C_{2}$) from both the real and simulated exposure-corrected skies, and its square root was taken as the measure of the quadrupole amplitude. The comparison, shown in \autoref{fig:15}, reveals that the observed quadrupole amplitude is only $0.28\sigma$ away from the mean of the exposure-convolved isotropic distribution, corresponding to a $p$-value of $0.393$. Hence, no statistically significant quadrupole anisotropy remains once the exposure correction is applied. 

These results jointly demonstrate that both the dipole and quadrupole anisotropies reported in the uncorrected BATSE GRB maps are artifacts of the instrument’s non-uniform exposure. Once this effect is rigorously modeled and incorporated into the isotropic simulations, the BATSE GRB sky becomes fully consistent with statistical isotropy. This emphasizes the crucial role of accounting for instrumental systematics in any large-scale anisotropy analysis based on GRB datasets.

Motivated by the exposure-corrected BATSE results discussed above, we now revisit the interpretation of the Fermi GBM quadrupole signal. For the Fermi GBM dataset, we find that while the dipole amplitude remains fully consistent with isotropy, the quadrupole amplitude exhibits a deviation at the $\sim3\sigma$ level when compared with isotropic simulations. Although such a level of statistical significance is not sufficient to establish a definitive departure from isotropy, it is commonly regarded as suggestive in cosmological analyses and therefore merits careful consideration. The BATSE dataset allows for a rigorous treatment of instrumental exposure through the availability of its well-characterized 5B exposure function, but no equivalent full-sky exposure maps are publicly available for the Fermi GBM instrument. Consequently, we were unable to perform a similar exposure-corrected isotropy analysis for the Fermi GRB dataset. The physical interpretation of the quadrupole excess in the Fermi data must therefore be treated with caution. 

In addition to instrumental exposure non-uniformities, several other observational effects may plausibly contribute to the quadrupole feature observed in the Fermi GBM GRB sky. The GBM instrument employs multiple detectors with different orientations and energy-dependent sensitivities, and its effective sky coverage varies with time due to orbital precession, Earth occultation, and passages through the South Atlantic Anomaly. These factors can introduce large-scale modulations in detection efficiency that are not strictly isotropic and may project preferentially onto low-order multipoles such as the quadrupole. Furthermore, the GBM trigger algorithm operates with background-dependent thresholds that vary with detector orientation, geomagnetic environment, and sky position, potentially introducing subtle selection biases. Even modest, percent-level variations in detection efficiency across broad sky regions could give rise to apparent quadrupole amplitudes of the magnitude observed here. In the absence of detailed, time-resolved exposure and trigger-efficiency maps for Fermi GBM, it is not currently possible to disentangle the relative contributions of these effects from a potential intrinsic signal. These considerations reinforce the interpretation of the Fermi quadrupole excess as a moderate-significance and potentially instrumental feature, warranting caution and motivating future analyses based on improved exposure characterization.

Although the analogy with the BATSE analysis where the apparent quadrupole anisotropy vanishes after accounting for instrumental exposure suggests that the Fermi quadrupole signal is likely influenced by residual exposure-related systematics, we cannot definitively rule out the presence of a weak intrinsic quadrupole component in the Fermi GRB distribution. Furthermore, since the observed Fermi dipole amplitude lies slightly below the mean of the isotropic distribution, the effect of a hypothetical exposure correction on the dipole and quadrupole significances cannot be reliably anticipated. The Fermi results thus remain intriguing, and the observed quadrupole excess should be regarded as a moderate-significance feature and a potential anomaly rather than conclusively attributed to either instrumental or cosmological origins.

\section{Conclusions}
\label{sec:conclusion}
In this work, we performed a detailed statistical investigation of the angular isotropy of gamma-ray bursts (GRBs) using two of the most comprehensive and widely studied  GRB catalogs to date: BATSE and Fermi GBM. Employing a statistical framework built on the HEALPix formalism, we quantified the large-scale angular distribution of GRBs and compared the observed dipole and quadrupole amplitudes with those expected from Monte Carlo realizations of isotropic skies. Our approach, which directly estimates the statistical significance of observed anisotropies by constructing empirical probability density functions, allows for a robust and model-independent assessment of isotropy across the celestial sphere.

The analysis of the BATSE dataset, containing 2702 GRBs, reveals no statistically significant dipole anisotropy, with the observed dipole amplitude lying well within the $1\sigma$ region of the isotropic distribution. The Fermi GBM dataset, consisting of 4032 GRBs, also shows no evidence of a preferred dipolar direction, supporting the hypothesis of large-scale isotropy in GRB distributions. However, when extending the analysis to the quadrupole term, both datasets exhibit signs of excess power. For BATSE, the quadrupole amplitude initially appears significant at the $3.7\sigma$ level, while for Fermi GBM, a milder but still notable $3\sigma$ excess is observed.

To investigate the impact of instrumental effects, we further applied exposure corrections to the BATSE data using its publicly available 5B exposure map. After accounting for non-uniform sky exposure, both the dipole and quadrupole signals are found to be statistically consistent with isotropy, suggesting that the earlier quadrupole excess was primarily instrumental in origin. Unfortunately, due to the absence of a publicly available full-sky exposure model for Fermi GBM, a similar correction could not be applied to that dataset. The analogy with BATSE suggests that the apparent Fermi quadrupole anisotropy is also likely to be an artifact of non-uniform sky exposure rather than a cosmological signal. However, the apparent quadrupole excess in the Fermi GBM dataset should be interpreted cautiously in the absence of an exposure-corrected analysis. Future GRB datasets with improved exposure characterization and redshift completeness will be essential for determining whether this feature persists as a genuine cosmological signal or is fully explained by observational effects.

When interpreting the statistical significances reported above, it is also important to note that they are primarily based on 500 Monte Carlo realizations of isotropic GRB skies. With this number of simulations, the minimum directly resolvable tail probability is of order $p \sim 2\times10^{-3}$ (\autoref{tab:dipole_significance}), implying that significance estimates approaching the $\sim3\sigma$ level are subject to additional uncertainty arising from finite sampling of the distribution tails. To mitigate this limitation, we construct smooth probability density functions using kernel density estimation (KDE), which reduces the impact of shot noise and sampling fluctuations in the extreme tails of the empirical distributions. We therefore adopt a deliberately conservative interpretation of results near the $\sim3\sigma$ level, treating them as moderate-significance and suggestive rather than definitive detections. Importantly, our main conclusions such as the disappearance of the BATSE quadrupole signal after exposure correction and the consistency of dipole amplitudes with isotropy do not rely on precise tail probabilities and remain robust against modest variations in the estimated significance. Larger simulation ensembles would be required for higher-precision tail characterization, but are not expected to qualitatively alter the conclusions presented here.

The robustness of our methodology was validated through controlled simulations of dipole-injected skies, demonstrating that the framework can reliably detect anisotropies as low as $a = 0.09$ for datasets comparable in size to BATSE. Moreover, tests with larger simulated samples containing $20000$ mock GRBs confirmed that the sensitivity improves significantly with increasing sample size, with detectable dipole amplitudes as small as $a = 0.022$. These results establish the reliability and precision of the adopted method for isotropy testing using current and future GRB catalogs. The strengths of our analysis lie in its simplicity, reproducibility, and empirical grounding. By relying on spherical harmonic decomposition and Monte Carlo simulations, we avoid assumptions tied to specific cosmological models or redshift distributions. Moreover, the use of two independent GRB catalogues adds confidence to the null result. However, some caveats remain. Our analysis is limited to angular positions and does not incorporate redshift information, which could offer deeper insight into the evolution of isotropy with cosmic time.

Our findings are in broad agreement with those that report no significant evidence for large-scale anisotropy in GRB distributions \citep{briggs96, bernui08, ripa19, andrade19}. In particular, our results align with works that found long-duration GRBs to be consistent with isotropy, and contribute additional support using an updated, simulation-backed dipole analysis. At the same time, our conclusions contrast with some earlier claims of anisotropy or clustering in GRB datasets \citep{mesazaros19, vavrek08, balazas99, tarnopolski17}, particularly among short-duration GRBs. These discrepancies highlight the importance of distinguishing between GRB subpopulations. Given their distinct astrophysical origins and environments, these two populations may, in principle, exhibit different large-scale anisotropy signatures, as suggested by several earlier studies. If the long- and short-duration GRB populations possess anisotropies of differing amplitudes or orientations, a combined analysis could partially suppress or mask such signals, yielding an apparently isotropic distribution. While separating the sample into sub-classes would provide a more physically refined test, the significantly reduced sample sizes particularly for short-duration GRBs and additional classification and selection uncertainties currently limit the statistical power of such an approach. Our results should therefore be interpreted as constraints on the isotropy of the combined GRB population, and future studies with larger, better-classified samples will be essential for probing anisotropy in individual GRB sub-populations.

Further, several studies have reported large-scale spatial features in the GRB distribution, such as the Giant GRB Ring and other clustering signals identified using redshift information and three-dimensional spatial analyses. It is important to clarify that these results probe a different regime of the GRB distribution than the present work. Our analysis is specifically designed to test large-scale angular isotropy through low-order spherical harmonic moments, which are sensitive to global, coherent anisotropies across the entire sky. In contrast, reported GRB structures are typically identified using localized clustering or pattern-recognition techniques in three-dimensional space and may be confined to specific redshift ranges or angular regions. Consequently, the angular scales and statistical methodologies involved are not directly comparable. The absence of significant dipole anisotropy and the consistency with isotropy at the quadrupole level once instrumental exposure effects are accounted for do not rule out the existence of such localized or redshift-dependent GRB structures. Rather, our results place constraints on the presence of large-scale, all-sky anisotropies in the GRB distribution, implying that any previously reported structures, if real, do not dominate the global angular distribution of GRBs or produce a detectable imprint in the lowest angular multipoles.

We also emphasize that the present study is restricted to the two-dimensional angular distribution of GRBs on the sky. While GRBs are intrinsically three-dimensional cosmological tracers, a fully spatial isotropy test or an analysis in redshift shells \citep{ferreira24} remains challenging with current data. Only a relatively small fraction of GRBs detected by all-sky instruments such as BATSE and Fermi GBM have reliably measured redshifts, and these subsamples are affected by strong selection effects arising from follow-up strategies, detector sensitivities, and sky coverage. Under these circumstances, angular analyses provide the most statistically robust and least biased framework for testing large-scale isotropy. The results presented here should therefore be regarded as a necessary and well-motivated first step. Future GRB catalogs with substantially larger and more homogeneous redshift coverage will be essential for extending these tests into three dimensions and for probing the isotropy of the Universe as a function of cosmic epoch.

Despite these limitations, the consistency of our results across two independent GRB catalogs, spanning different missions and observational epochs, provides strong support for the statistical isotropy of the Universe on the largest angular scales accessible with current GRB data. Our analysis demonstrates that, once instrumental exposure effects are properly accounted for, neither dipole nor quadrupole anisotropies persist in the BATSE GRB sky, and that the Fermi GBM results are consistent with this picture within present observational constraints. As GRB samples continue to grow and redshift completeness improves through sustained multi-wavelength follow-up and future high-energy missions, the methodology presented here can be naturally extended to redshift-resolved analyses. Such developments will enable increasingly stringent tests of the Cosmological Principle and further clarify the role of GRBs as powerful probes of the large-scale structure and evolution of the Universe.

\section{Acknowledgements}
DM thanks Jon Hakkila for sharing the BATSE 5B exposure function. DM would also like to thank P. N. Bhat for useful discussions regarding the FERMI GBM data. BP would like to acknowledge IUCAA, Pune for providing support through associateship programme. AM thanks UGC, Government of India for support through a Junior Research Fellowship.

\section{Data availability}
The GRB datasets used in this study are publicly avaialble at \url{https://heasarc.gsfc.nasa.gov/W3Browse/cgro/batsegrb.html} and \url{https://heasarc.gsfc.nasa.gov/W3Browse/fermi/fermigbrst.html}. Additional data generated and analyzed during this study are available from the authors upon reasonable request.

%\bibliographystyle{unsrt}
%\bibliography{references}

\appendix

\section{Appendix: Stability tests with different HEALPix resolutions}
\label{appendix:nsides}

In this Appendix, we examine the stability and robustness of our results against changes in the HEALPix resolution parameter, \(N_{\mathrm{side}}\). Since the estimation of low-order multipoles such as dipole and quadrupole depends on how the celestial sphere is pixelized, it is important to confirm that our conclusions are not influenced by the chosen resolution. The following analysis tests the consistency of our dipole and quadrupole results for the BATSE dataset when different values of \(N_{\mathrm{side}}\) are adopted.

\subsection{Effects of HEALPix resolution parameter on dipole and quadrupole estimates for the exposure-corrected BATSE data}

The HEALPix resolution parameter \(N_{\mathrm{side}}\) determines the total number of equal-area pixels on the celestial sphere, given by \(N_{\mathrm{pix}} = 12 \times N_{\mathrm{side}}^{2}\). A higher \(N_{\mathrm{side}}\) value provides finer angular sampling but may increase statistical noise for sparse datasets, whereas lower values smooth the map and can slightly  affect the estimation of large-scale moments such as dipole and quadrupole amplitudes. Hence, testing the sensitivity of our results to \(N_{\mathrm{side}}\) provides an important consistency check for the robustness of the analysis.

To evaluate this, we repeated the dipole and quadrupole analyses using exposure-convolved isotropic skies and the BATSE dataset for two higher resolutions, \(N_{\mathrm{side}} = 32\) and \(N_{\mathrm{side}} = 64\), in addition to our baseline \(N_{\mathrm{side}} = 8\). The exposure correction procedure, described in Section~\ref{sec:batse_exposure_correction}, was consistently applied to all realizations to isolate the effects of pixelization.

\begin{figure*}[htbp!]
\centering\includegraphics[width=12cm]{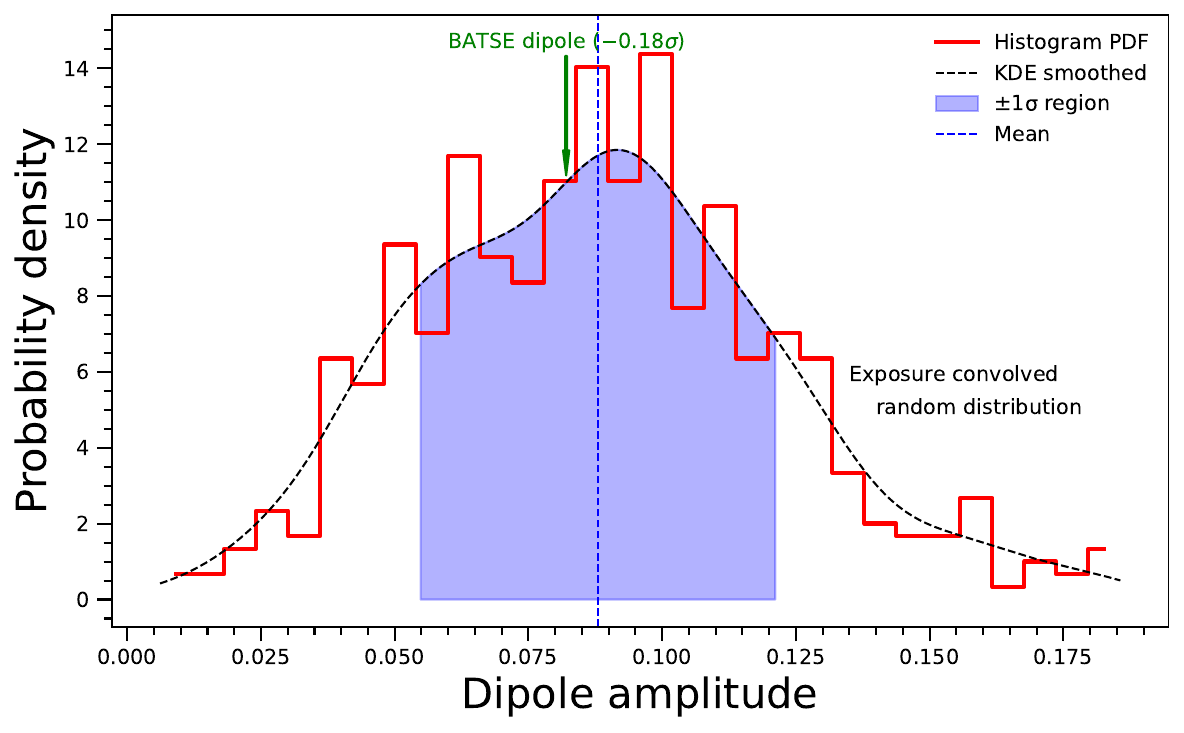}
\caption{This compares the observed dipole amplitude for BATSE data (vertical downward arrow) with the estimated PDF obtained from Monte Carlo simulations of exposure-convolved isotropic skies containing 2702 points each. The position of the observed dipole amplitude is $-0.18\sigma$ away from the mean of the isotropic distribution. HEALPix resolution parameter \(N_{\mathrm{side}}=32\) has been used.}
\label{fig:13}
\end{figure*}

\begin{figure*}[htbp!]
\centering\includegraphics[width=12cm]{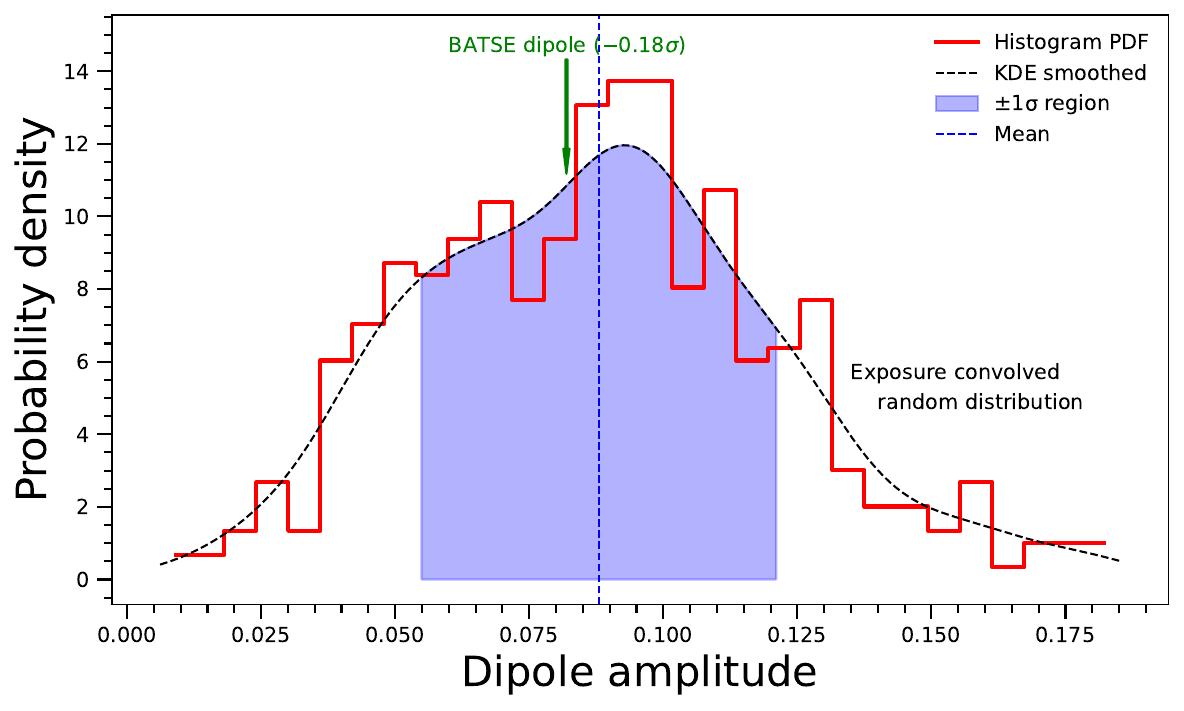}
\caption{Same as Figure~\ref{fig:13}, but with HEALPix resolution parameter \(N_{\mathrm{side}}=64\). The observed dipole amplitude is $-0.18\sigma$ away from the mean of the isotropic distribution.}
\label{fig:14}
\end{figure*}

Figures~\ref{fig:13} and~\ref{fig:14} present the resulting PDFs for the dipole amplitudes. At both resolutions, the observed BATSE dipole amplitude lies extremely close to the mean of the isotropic distribution, deviating by only $-0.18\sigma$. This confirms that the dipole estimate is highly stable and statistically insignificant across all pixelization scales, reaffirming that the exposure-corrected BATSE sky remains consistent with isotropy.

\begin{figure*}[htbp!]
\centering\includegraphics[width=12cm]{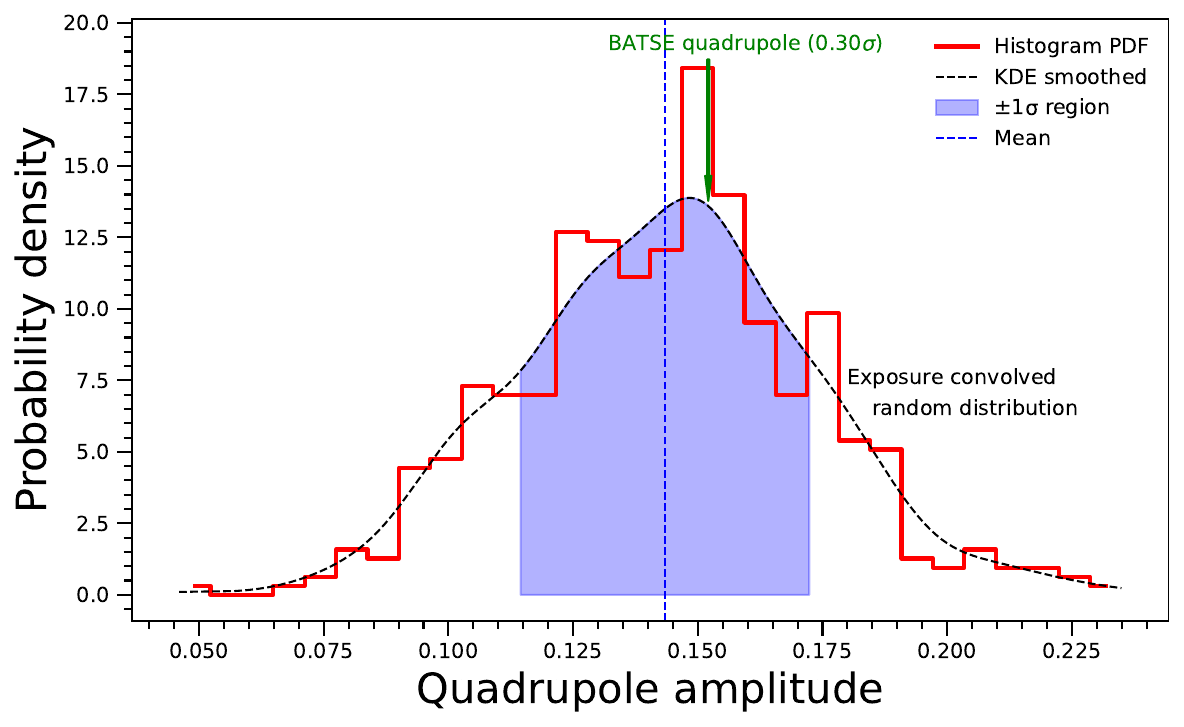}
\caption{This compares the observed quadrupole amplitude for BATSE data (vertical downward arrow) with the estimated PDF obtained from Monte Carlo simulations of exposure-convolved isotropic skies containing 2702 points each. The position of the observed quadrupole amplitude is $0.30\sigma$ away from the mean of the isotropic distribution. HEALPix resolution parameter \(N_{\mathrm{side}}=32\) has been used.}
\label{fig:16}
\end{figure*}

\begin{figure*}[htbp!]
\centering\includegraphics[width=12cm]{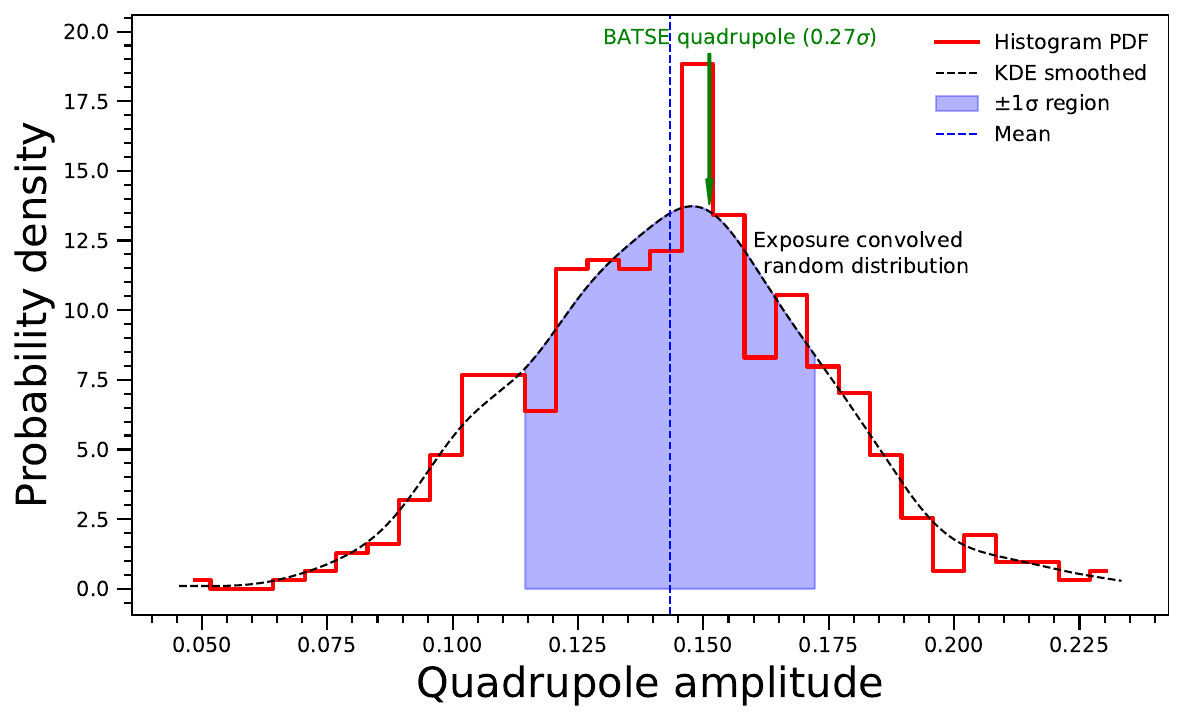}
\caption{Same as Figure~\ref{fig:16}, but with HEALPix resolution parameter \(N_{\mathrm{side}}=64\). The observed quadrupole amplitude is $0.27\sigma$ away from the mean of the isotropic distribution.}
\label{fig:17}
\end{figure*}

A similar level of consistency is observed for the quadrupole analysis, as shown in Figures~\ref{fig:16} and~\ref{fig:17}. At \(N_{\mathrm{side}} = 32\), the quadrupole amplitude differs from the isotropic mean by $0.30\sigma$, while at \(N_{\mathrm{side}} = 64\) the deviation is $0.27\sigma$. These small deviations confirm that the quadrupole results, like the dipole, are insensitive to the choice of HEALPix resolution.

Overall, this consistency check demonstrates that our conclusions regarding the isotropy of the BATSE GRB sky are robust to variations in the pixelization scale. The observed dipole and quadrupole amplitudes are not artifacts of map resolution, but reflect the intrinsic statistical properties of the data after appropriate exposure correction.

\subsection{Stability of dipole and quadrupole estimates for raw BATSE and Fermi GRB data}
\label{app:raw_nsides}

In addition to the stability tests performed for the exposure-corrected BATSE dataset, we also examine the robustness of the dipole and quadrupole estimates obtained from the raw, uncorrected GRB sky maps. This analysis is motivated by the need to verify that the elevated quadrupole amplitudes identified in the raw BATSE and Fermi datasets are not artifacts of the chosen HEALPix pixelization scale, but instead represent stable features of the uncorrected sky distributions.

Following the same procedure adopted in the main analysis, we recompute the dipole and quadrupole amplitudes for the raw BATSE and Fermi GRB skies using higher HEALPix resolutions, specifically \(N_{\mathrm{side}} = 32\) and \(N_{\mathrm{side}} = 64\), in addition to the baseline resolution \(N_{\mathrm{side}} = 8\). For each resolution, the observed multipole amplitudes are compared with probability density functions constructed from 500 Monte Carlo realizations of isotropic skies containing the same number of sources as the corresponding datasets.

\begin{figure*}[htbp!]
\centering\includegraphics[width=12cm]{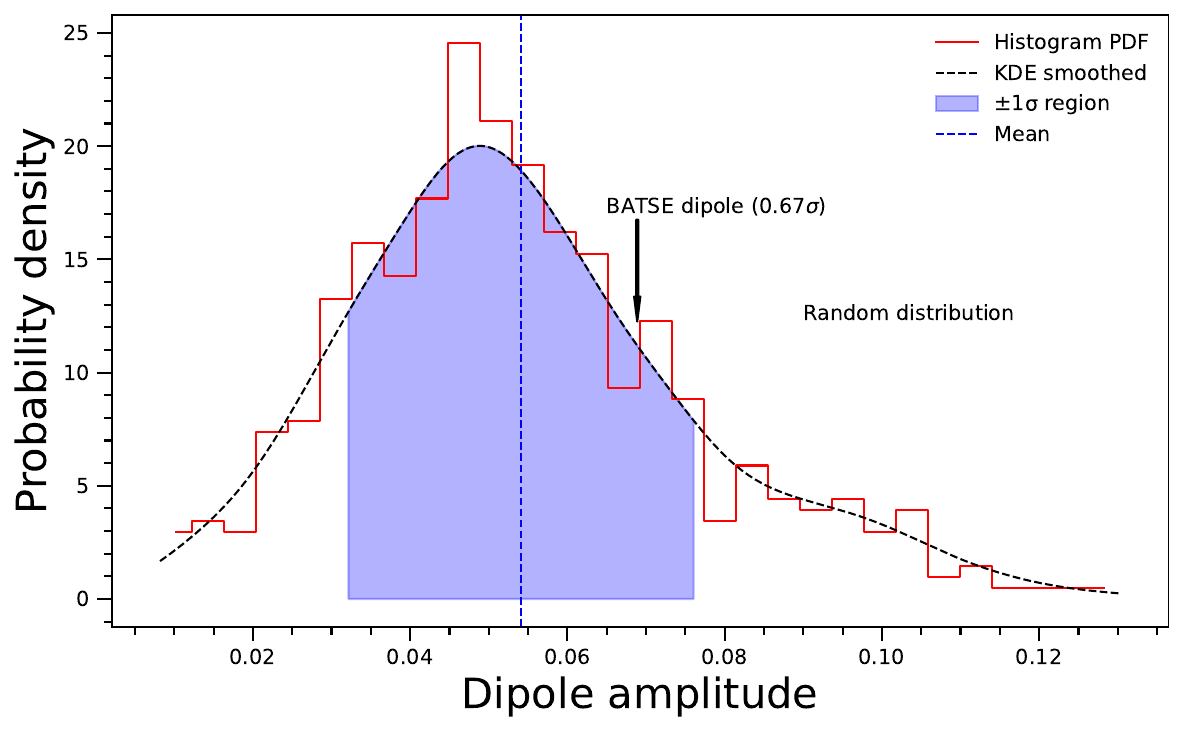}
\caption{Comparison of the observed dipole amplitude for the raw BATSE dataset (vertical downward arrow) with the estimated PDF obtained from Monte Carlo simulations of isotropic skies containing 2702 points each. The position of the observed dipole amplitude lies within the isotropic distribution and shows no statistically significant deviation. In this comparison, HEALPix resolution parameter \(N_{\mathrm{side}} = 32\) has been used.}
\label{fig:18}
\end{figure*}

\begin{figure*}[htbp!]
\centering\includegraphics[width=12cm]{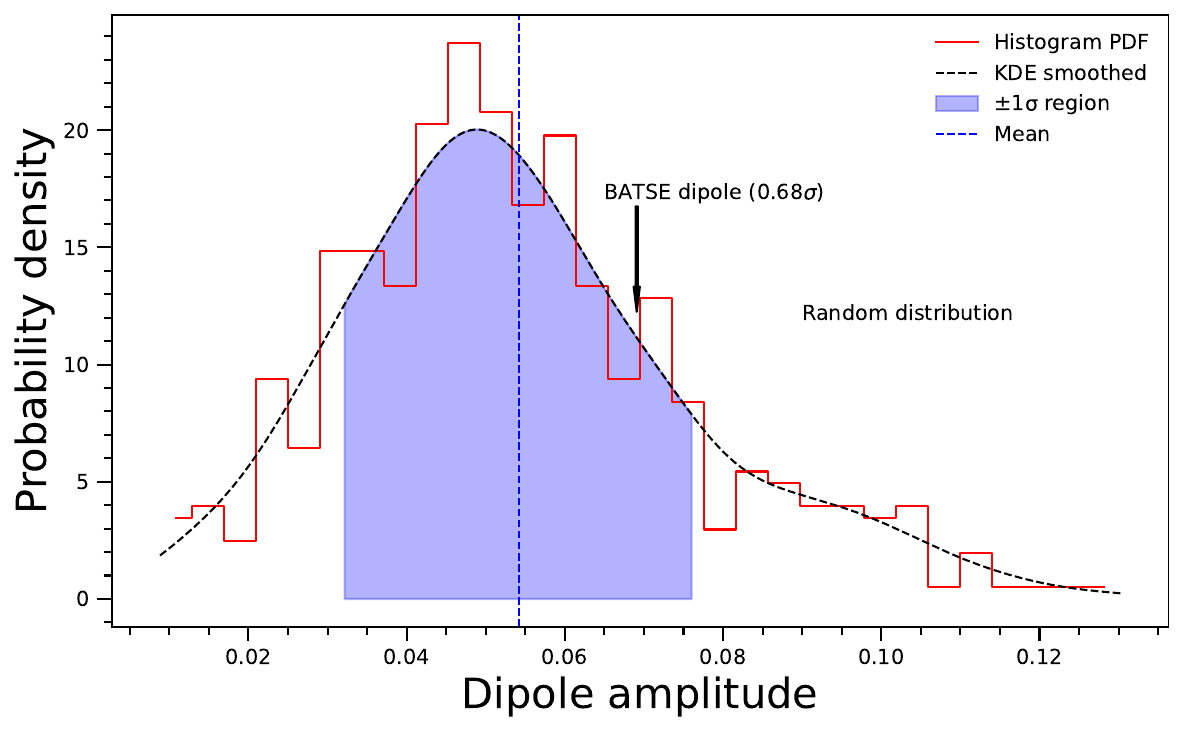}
\caption{Same as Figure~\ref{fig:18}, but with HEALPix resolution parameter \(N_{\mathrm{side}} = 64\). The observed dipole amplitude remains consistent with statistical isotropy, indicating stability against changes in pixelization scale.}
\label{fig:19}
\end{figure*}

\begin{figure*}[htbp!]
\centering\includegraphics[width=12cm]{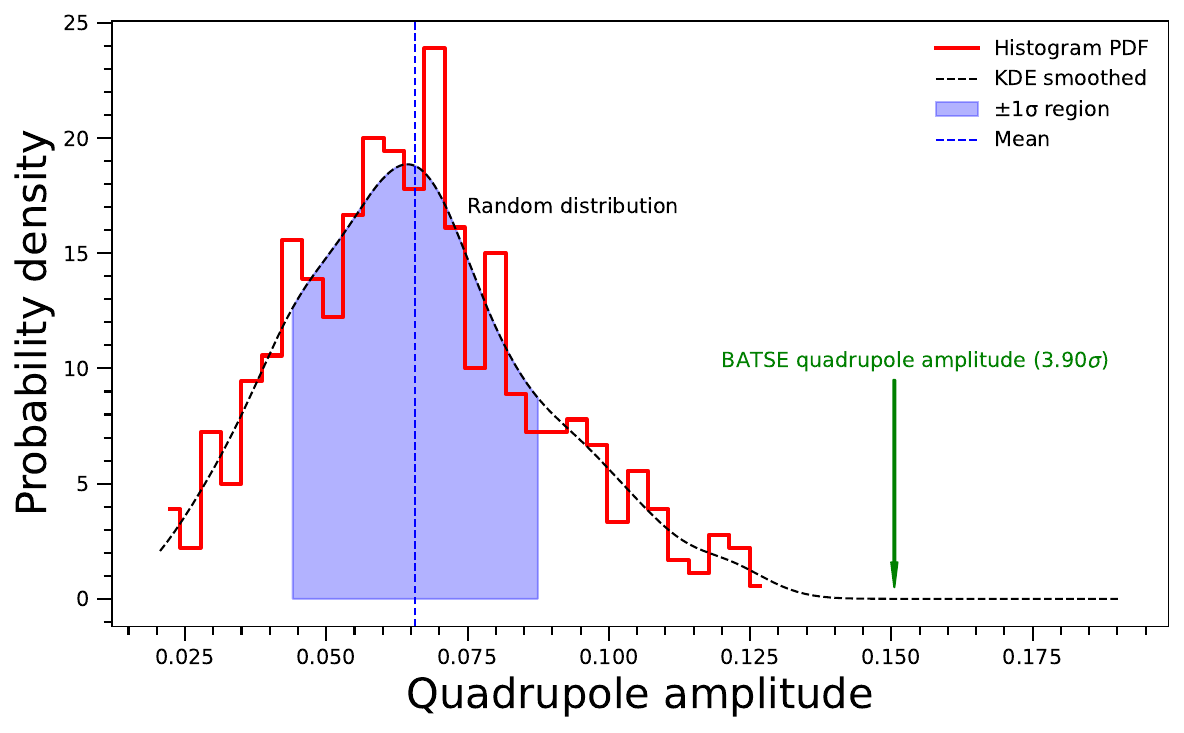}
\caption{Comparison of the observed quadrupole amplitude for the raw BATSE dataset (vertical downward arrow) with the estimated PDF obtained from Monte Carlo simulations of isotropic skies containing 2702 points each. The observed quadrupole amplitude remains elevated relative to the isotropic mean at a significance close to the $\sim3.9\sigma$ level. In this comparison, HEALPix resolution parameter \(N_{\mathrm{side}} = 32\) has been used.}
\label{fig:20}
\end{figure*}

\begin{figure*}[htbp!]
\centering\includegraphics[width=12cm]{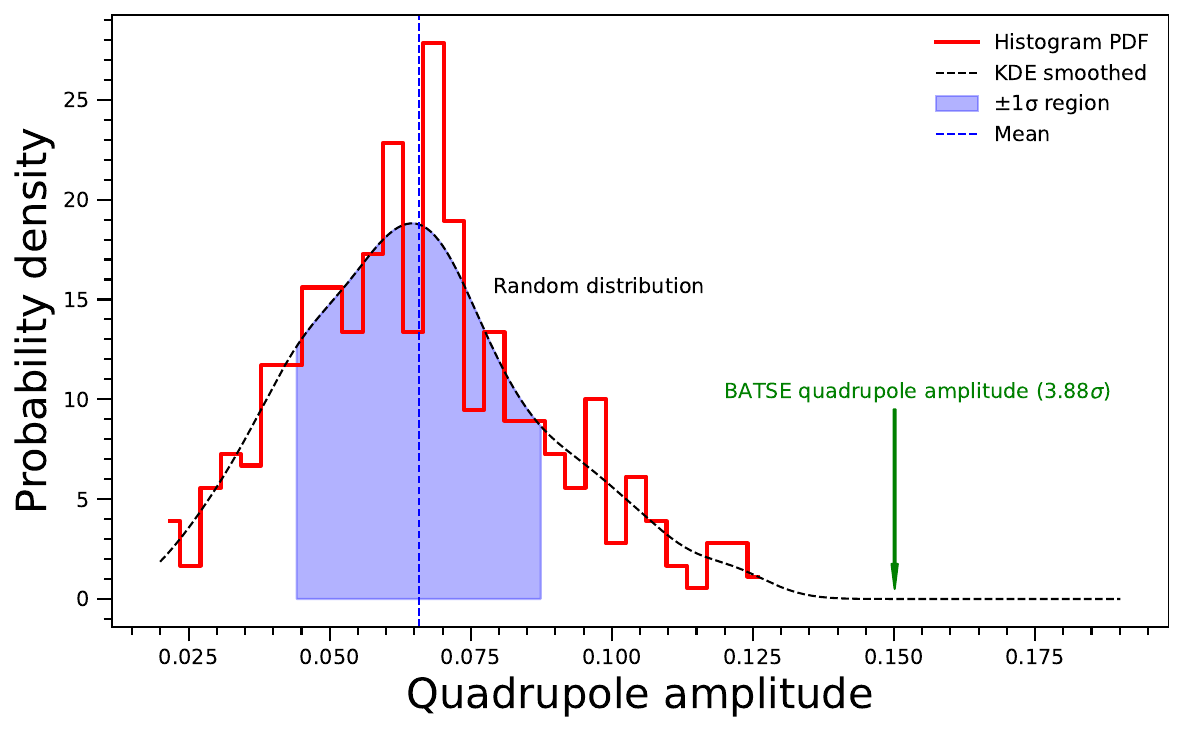}
\caption{Same as Figure~\ref{fig:20}, but with HEALPix resolution parameter \(N_{\mathrm{side}} = 64\). The persistence of the elevated quadrupole amplitude confirms that the signal is not an artifact of the chosen pixelization scale.}
\label{fig:21}
\end{figure*}

For the raw BATSE data, the quadrupole amplitude remains consistently elevated relative to the isotropic expectation across all tested resolutions, with the observed deviation remaining close to the $\sim3.9\sigma$ level (Figures~\ref{fig:20} and~\ref{fig:21}). Similarly, the dipole amplitude continues to lie well within the isotropic distribution, independent of the chosen pixelization scale (Figures~\ref{fig:18} and~\ref{fig:19}). This confirms that the apparent quadrupole excess observed in the raw BATSE sky is not driven by the choice of \(N_{\mathrm{side}}\), but reflects a stable property of the uncorrected map.

\begin{figure*}[htbp!]
\centering\includegraphics[width=12cm]{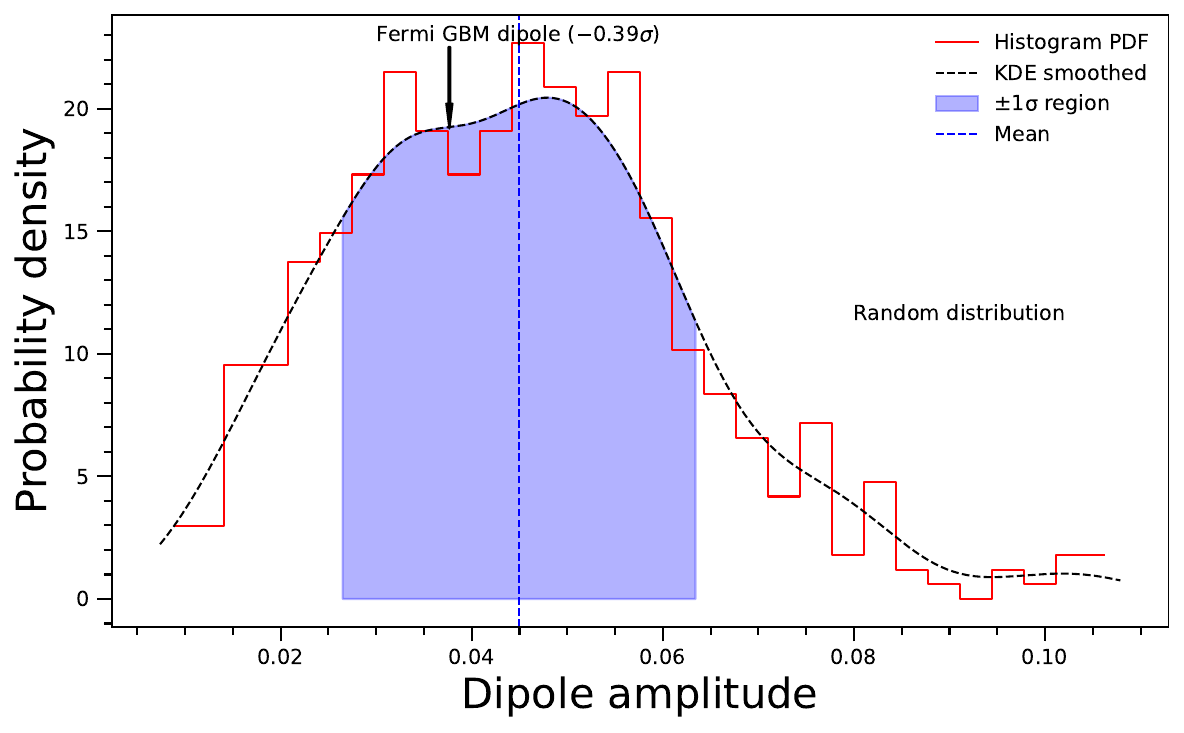}
\caption{Comparison of the observed dipole amplitude for the raw Fermi GBM dataset (vertical downward arrow) with the estimated PDF obtained from Monte Carlo simulations of isotropic skies containing 4032 points each. The observed dipole amplitude lies well within the isotropic distribution. In this comparison, HEALPix resolution parameter \(N_{\mathrm{side}} = 32\) has been used.}
\label{fig:22}
\end{figure*}

\begin{figure*}[htbp!]
\centering\includegraphics[width=12cm]{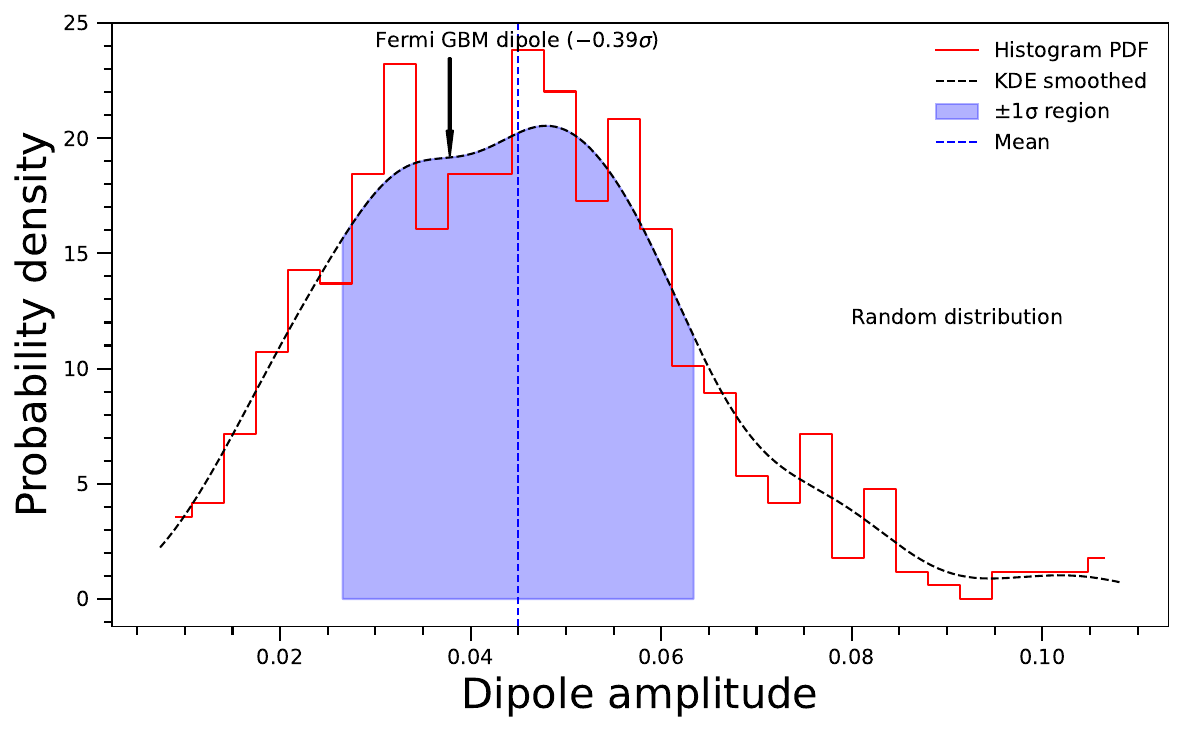}
\caption{Same as Figure~\ref{fig:22}, but with HEALPix resolution parameter \(N_{\mathrm{side}} = 64\). The dipole amplitude remains statistically consistent with isotropy, demonstrating robustness to changes in HEALPix resolution.}
\label{fig:23}
\end{figure*}

\begin{figure*}[htbp!]
\centering\includegraphics[width=12cm]{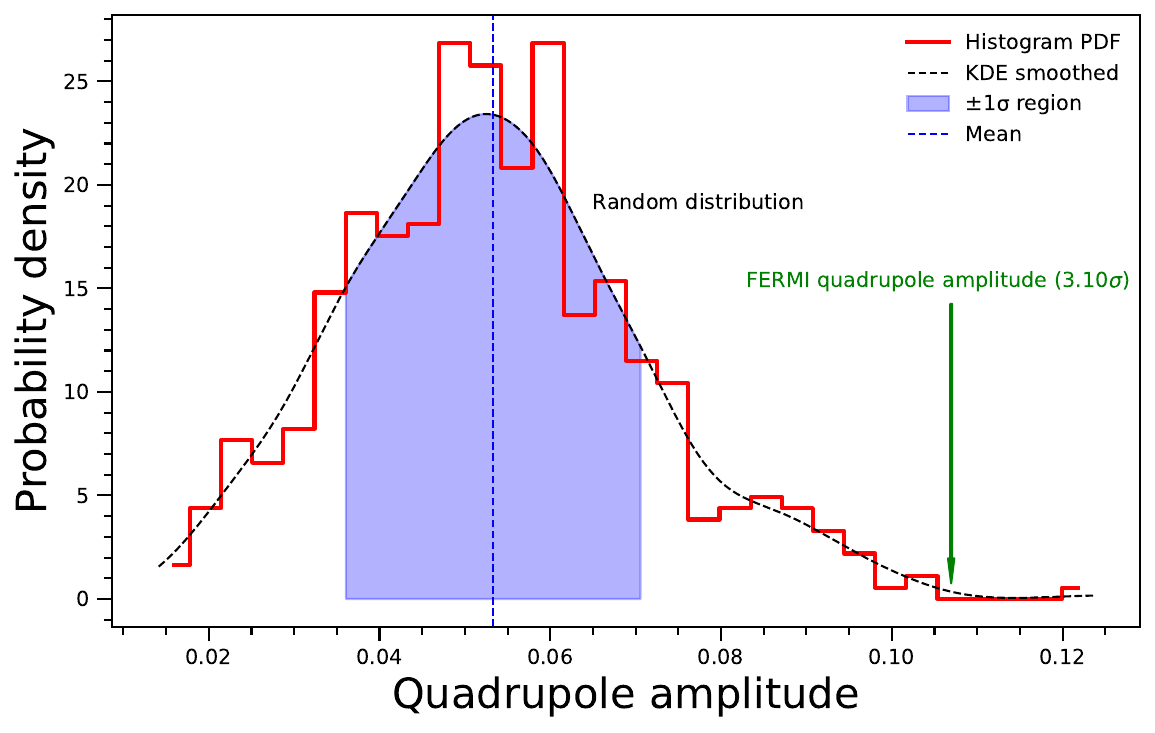}
\caption{Comparison of the observed quadrupole amplitude for the raw Fermi GBM dataset (vertical downward arrow) with the estimated PDF obtained from Monte Carlo simulations of isotropic skies containing 4032 points each. The observed quadrupole amplitude shows a deviation at the $\sim3\sigma$ level relative to the isotropic mean. In this comparison, HEALPix resolution parameter \(N_{\mathrm{side}} = 32\) has been used.}
\label{fig:24}
\end{figure*}

\begin{figure*}[htbp!]
\centering\includegraphics[width=12cm]{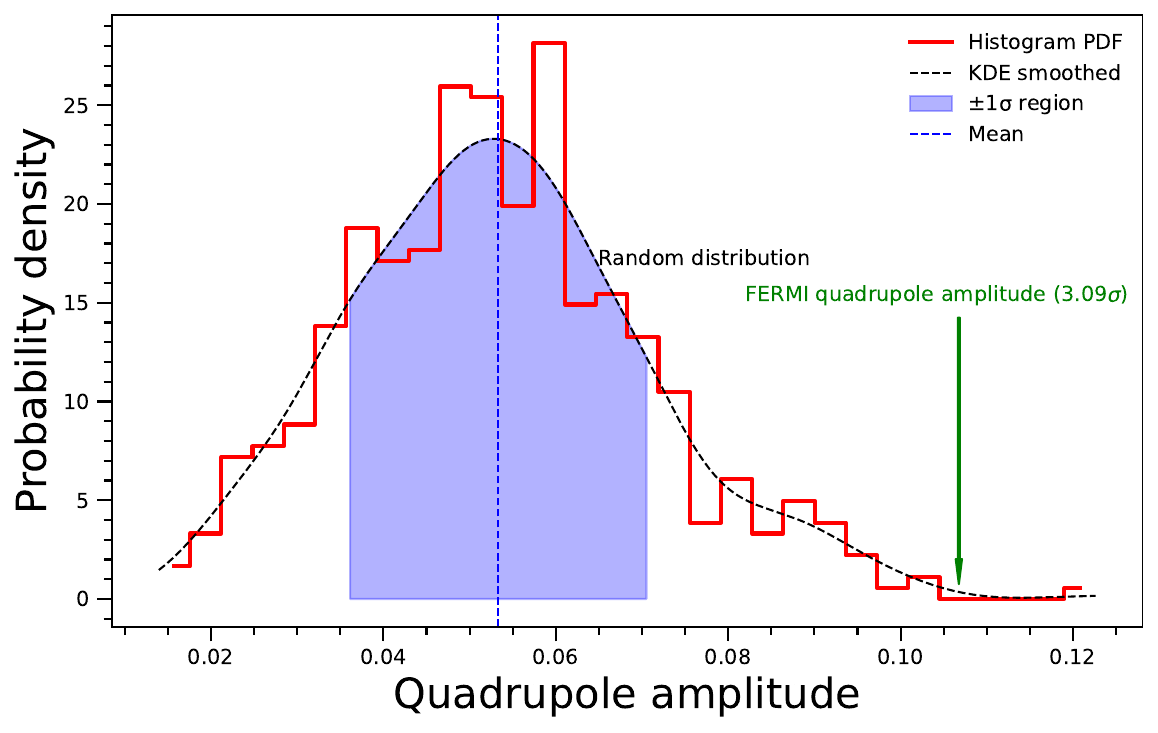}
\caption{Same as Figure~\ref{fig:24}, but with HEALPix resolution parameter \(N_{\mathrm{side}} = 64\). The quadrupole excess persists across resolutions, indicating that it is not driven by the choice of pixelization scale.}
\label{fig:25}
\end{figure*}

An analogous behavior is found for the raw Fermi GBM dataset. The quadrupole amplitude remains systematically offset from the isotropic mean at the $\sim3\sigma$ level for both \(N_{\mathrm{side}} = 32\) and \(N_{\mathrm{side}} = 64\) (Figures~\ref{fig:24} and~\ref{fig:25}), while the dipole amplitude shows no statistically significant deviation from isotropy at any resolution (Figures~\ref{fig:22} and~\ref{fig:23}). These results demonstrate that the quadrupole excess identified in the raw Fermi sky is likewise robust to reasonable variations in HEALPix resolution.

Taken together, these stability tests confirm that the moderate-significance quadrupole signals observed in the raw BATSE and Fermi GRB datasets are not artifacts of pixelization. While the exposure-corrected BATSE analysis shows that instrumental effects can fully account for this signal, particularly in that case, the persistence of the quadrupole excess across resolutions reinforces the need for careful exposure modeling particularly for Fermi GBM before definitive physical interpretations can be drawn.

\end{document}